\documentclass[fleqn,usenatbib]{mnras}

\usepackage{newtxtext,newtxmath}

\usepackage[T1]{fontenc}

\DeclareRobustCommand{\VAN}[3]{#2}
\let\VANthebibliography\thebibliography
\def\thebibliography{\DeclareRobustCommand{\VAN}[3]{##3}\VANthebibliography}

\usepackage{graphicx}	
\usepackage{amsmath}	
\usepackage{booktabs}
\usepackage[T1]{fontenc}
\usepackage{icomma}

\usepackage[flushleft]{threeparttable}
\usepackage{hyperref}

\newcommand{\waspb}{WASP-127\,b}
\newcommand{\wasp}{WASP-127}
\newcommand{\water}{H$_2$O}
\newcommand{\methane}{CH$_4$}
\newcommand{\diox}{CO$_2$}
\newcommand{\tp}{\ensuremath{T_{\rm P}}}
\newcommand{\kp}{$K_{\rm P}$}
\newcommand{\vsys}{$v_{\rm sys}$}
\newcommand{\vvv}[1]{$v_{\rm #1}$}
\newcommand{\vmr}{\ensuremath{\text{VMR}[{\rm H}_2{\rm O}]}}
\newcommand{\cloud}{$P_{\rm cloud}$}
\newcommand{\logl}{$\ln \mathcal{L}$}
\newcommand{\kms}{km\,s$^{-1}$}
\newcommand{\um}{$\mu$m}

\newcommand{\inc}[2]{^{+#1}_{-#2}}
\newcommand{\expref}[1]{$^{\,\textrm{#1}}$}

\newcommand{\pcun}{2}
\newcommand{\pcdeux}{2}
\newcommand{\pctrois}{5}

\newcommand{\vpeakJR}{$-0.3$}
\newcommand{\vpeakJRerr}{$\inc{0.7}{0.7}$}
\newcommand{\fwhmJR}{$9.1\inc{0.5}{0.5}$}
\newcommand{\cloudfracJR}{$0.49\inc{0.09}{0.10}$}

\newcommand{\waterJR}{$-3.0$}
\newcommand{\waterJRerr}{$\inc{0.5}{0.6}$}
\newcommand{\dioxJR}{$-3.7$}
\newcommand{\dioxJRerr}{$\inc{0.8}{0.6}$}
\newcommand{\OHJR}{$-3.0$}
\newcommand{\OHJRerr}{$\inc{0.9}{0.7}$}
\newcommand{\NaJR}{$-6.6$}
\newcommand{\NaJRerr}{$\inc{2.0}{1.9}$}

\newcommand{\FeHLRR}{-5.5}
\newcommand{\FeHLRRerr}{$\inc{1.4}{3.5}$}
\newcommand{\COJR}{$-4.0$}
\newcommand{\FeHJR}{-8.7}

\newcommand{\waterdiox}{$0.8$}
\newcommand{\waterdioxerr}{$\inc{0.3}{0.5}$}

\newcommand{\CsurOdistnoOH}{$0.10$}
\newcommand{\CsurOdisterrnoOH}{$\inc{0.10}{0.06}$}

\newcommand{\CsurOupperlimtwosig}{$0.30$}

\newcommand{\sigttest}{$5.3$} 
\newcommand{\kpttestmax}{$148$}
\newcommand{\kpttestmaxerr}{$\inc{33}{57}$}
\newcommand{\vradttestmax}{$-6.8$}
\newcommand{\vradttestmaxerr}{$\inc{0.8}{0.8}$}
\newcommand{\vpeakttestkpo}{$-7.0$}
\newcommand{\vpeakttestkpoerr}{$\inc{0.5}{0.5}$}

\newcommand{\OHttest}{$3.0$} 
\newcommand{\OHv}{$-7.0$}

\usepackage{fancyhdr}

\defcitealias{Spake2021}{S21}

\title[High Resolution Transit spectroscopy of WASP-127\,\rm{b}]{CO or no CO? Narrowing the CO abundance constraint and recovering the H$_2$O detection in the atmosphere of WASP-127\,b using SPIRou}

\author[A. Boucher et al.]{Anne Boucher$^{1}$\thanks{E-mail: anne.boucher.3@umontreal.ca}, 
David Lafrenière$^{1}$,
Stefan Pelletier $^{1}$,
Antoine Darveau-Bernier$^{1}$, 
Michael Radica$^{1}$,
\newauthor Romain Allart$^{1,*}$,
Étienne Artigau$^{1,2}$,
Neil J. Cook$^{1,2}$,
Florian Debras$^{3}$,
René Doyon$^{1,2}$,
Eric Gaidos$^{4}$,
\newauthor Björn Benneke$^{1}$,
Charles Cadieux$^{1}$,
Andres Carmona$^{5}$,
Ryan Cloutier$^{6,†}$,
Pía Cortés-Zuleta$^{7}$,
\newauthor Nicolas B. Cowan$^{8,9}$,
Xavier Delfosse$^{5}$,
Jean-François Donati$^{3}$,
Pascal Fouqué$^{3,10}$,
Thierry Forveille$^{5}$,
\newauthor Konstantin Grankin$^{11}$,
 Guillaume Hébrard$^{12}$,
Jorge H. C. Martins$^{13}$,
Eder Martioli$^{12,14}$,
Adrien Masson$^{15}$,
\newauthor and Sandrine Vinatier$^{15}$
\\
$^{1}$ Institut Trottier de Recherche sur les Exoplanètes, Université de Montréal, Département de Physique, C.P. 6128 Succ. Centre-ville, Montréal, QC H3C 3J7, Canada \\
$^{2}$ Observatoire du Mont-Mégantic, Université de Montréal, Notre-Dame-des-Bois, J0B 2E0, Canada \\
$^{3}$ Université de Toulouse, CNRS, IRAP, 14 av. Belin, 31400 Toulouse, France \\
$^{4}$ Department of Earth Sciences, University of Hawai’i at Manoa, Honolulu, HI 96822 USA \\
$^{5}$ Université Grenoble Alpes, CNRS, IPAG, 38000 Grenoble, France \\
$^{6}$ Center for Astrophysics | Harvard \& Smithsonian, 60 Garden Street, Cambridge, MA, 02138, USA \\
$^{7}$ Aix Marseille Univ, CNRS, CNES, LAM, Marseille, France \\
$^{8}$ Department of Earth \& Planetary Sciences, McGill University, 3450 rue University, Montréal, QC H3A 0E8, Canada  \\
$^{9}$ Department of Physics, McGill University, 3600 rue University, Montréal, QC H3A 2T8, Canada \\
$^{10}$ CFHT Corporation; 65-1238 Mamalahoa Highway, Kamuela, HI 96743, USA \\
$^{11}$ Crimean Astrophysical Observatory, Department of Stellar Physics, Nauchny, 298409, Crimea \\
$^{12}$ Institut d’astrophysique de Paris, UMR7095 CNRS, Sorbonne Université, 98 bis bd Arago, 75014 Paris, France  \\
$^{13}$ Instituto de Astrofísica e Ciências do Espaço, Universidade do Porto, CAUP, Rua das Estrelas, 4150-762 Porto, Portugal  \\
$^{14}$ Laboratório Nacional de Astrofísica, Rua Estados Unidos 154, Itajubá, MG 37504-364, Brazil \\
$^{15}$ LESIA, Observatoire de Paris, Université PSL, CNRS, Sorbonne Université, Université de Paris, 5 place Jules Janssen, 92195 Meudon, France
}

\date{Accepted XXX. Received YYY; in original form ZZZ}

\pubyear{2023}

\begin{document}
\label{firstpage}
\pagerange{\pageref{firstpage}--\pageref{lastpage}}
\maketitle

\begin{abstract}
Precise measurements of chemical abundances in planetary atmospheres are necessary to constrain the formation histories of exoplanets.  
A recent study of \waspb, a close-in puffy sub-Saturn orbiting its solar-type host star in 4.2\,d, using HST and \emph{Spitzer} revealed a feature-rich transmission spectrum with strong excess absorption at 4.5\,\um.
However, the limited spectral resolution and coverage of these instruments could not distinguish between CO and/or \diox\ absorption causing this signal, with both low and high C/O ratio scenarios being possible. 
Here we present near-infrared (0.9--2.5\,\um) transit observations of WASP-127\,b using the high-resolution SPIRou spectrograph, with the goal to disentangle CO from \diox\ through the 2.3\,\um\ CO band.  
With SPIRou, we detect \water\ at a $t$-test significance of \sigttest\,$\sigma$ and observe a tentative ($3\,\sigma$) signal consistent with OH absorption.
From a joint SPIRou + HST + \emph{Spitzer} retrieval analysis, we rule out a CO-rich scenario by placing an upper limit on the CO abundance of $\log_{10}$[CO] $<$\COJR, and estimate a $\log_{10}$[\diox]\ of \dioxJR\dioxJRerr, which is the level needed to match the excess absorption seen at 4.5\,\um.
We also set abundance constraints on other major C-, O-, and N-bearing molecules, with our results favoring low C/O (\CsurOdistnoOH\CsurOdisterrnoOH), disequilibrium chemistry scenarios.
We further discuss the implications of our results in the context of planet formation.
Additional observations at high and low-resolution will be needed to confirm these results and better our understanding of this unusual world.
\end{abstract}



\section{Introduction}
\label{sec:Intro}

The atmospheric metallicity and carbon-to-oxygen ratios (C/O) of exoplanets are tracers of their atmospheric chemistry, their formation location, and their migration history \citep{Madhusudhan2014, madhusudhan_exoplanetary_2019}. This stems from the fact that different volatile compounds (i.e. \water, \diox, \methane, CO, NH$_3$, etc.) condense at varying temperatures, and thus at different distances from the host star in the natal proto-planetary disk. This then introduces variations in the C/O ratio of the gaseous and condensed phases of matter which accrete onto a forming planet \citep{Oberg2011}. 
Therefore, the measurement of chemical abundances is a powerful tool to explore the origins and evolution pathways of exoplanets. 
Hot giant exoplanets with extended atmospheres are prime targets for these measurements thanks to their large sizes and scale heights relative to their host star, and high atmospheric temperature that keep most carbon and oxygen bearing molecules in vapour phase. 
This facilitates the remote detection of atomic and molecular species via transmission or emission spectroscopy. 
However, precise measurement of the C/O ratio remains a challenging task as it necessitates simultaneous constraints of all major C- and O-bearing molecules contained in the planet's atmosphere and an appropriate understanding of the chemistry \citep[e.g.,][]{Tsai2021_comparative}. The \textit{Hubble Space Telescope} Wide Field Camera 3 (HST/WFC3) G141 grism, widely used to study exoplanetary atmospheres \citep[e.g.,][and references therein]{Sing2016, welbanks_massmetallicity_2019, Pinhas2019}, provides access to the water-band feature at 1.4\,\um, but not to CO or \diox\ --- the bulk of whose spectral features lie further into the infrared. Complementary information could also be extracted from \emph{Spitzer} photometry, but this is limited to a few wide band-pass photometric data points --- rendering impossible the disentanglement of contributions from CO and \diox. Addressing this fundamental challenge requires additional observations which expand the limited wavelength range and spectral resolution of these instruments.

A first attempt at measuring the C/O by combining space- and ground-based observations was performed by \cite{madhusudhan_carbon-rich_2011}, where they inferred a carbon-rich atmosphere (C/O $\geq$ 1) in the day-side of the ultra-hot Jupiter WASP-12\,b. This sparked a debate, wherein some contested this result \citep[e.g.,][]{Cowan2012, crossfield_re-evaluating_2012, swain_probing_2013, Line2014, Kreidberg2015_wasp12}, while others further affirmed it \citep[e.g.,][]{fohring_ultracam_2013, stevenson_deciphering_2014}. Since then, analyses of multiple transit observations for other planets have been carried out to measure their C/O ratios. Most retrieved values have been closer to solar (C/O=0.54) than to 1 \citep{Line2014, Benneke2015}. Due to the large uncertainties on many C/O measurements, it is also unclear whether the general trend, if there is any, points toward super-\textit{stellar} values (i.e., when compared to host stars as opposed to the Sun), or indeed whether planetary C/O ratios are consistent with those of their host stars \citep{brewer_co_2017}. More recently, studies using ground-based high-resolution cross-correlation spectroscopy, with 
the combined constraints on \water\ and CO, showed evidence that certain hot Jupiters (HD~209458\,b, HD~189733\,b, Tau Bo\"otis\,b) have elevated C/O ratios \citep{Brogi2019, gandhi_hydra-h_2019, Giacobbe2021, Pelletier2021, Boucher2021} while others like WASP-77\,A\,b have C/O ratios closer to solar~\citep{Line2021}. 
As these various inferred atmospheric metallicities and C/O ratios could imply different formation conditions, more in depth measurements could reveal population trends on how and where in the protoplanetary disk giant planets form.

Here, we aim to measure the elemental abundances of the major molecules in the atmosphere of \waspb\ to gain insight into its formation and migration history. We use ground-based high-dispersion spectroscopy (HDS) in the near-infrared (NIR), which has proven to be a powerful probe of the composition of exoplanetary atmospheres  \citep[e.g.,][]{Snellen2010, Birkby2013, Nugroho2017, Brogi2018, Alonso2019_water, Boucher2021, Line2021}.


This paper is organized as follows: 
the next section is an overview of existing studies of \waspb\ and the remaining open science questions regarding this unusual system. In Section~\ref{sec:Obs_a2}, we present the observational setup and data. In Section~\ref{sec:analysis_a2} we briefly describe the data reduction, the telluric and stellar signal removal procedures. Section~\ref{sec:correl_a2} details the atmospheric modeling as well as the planetary signal extraction methods, and Section~\ref{sec:results_a2} presents the associated results. In Section~\ref{sec:Discuss_a2} we discuss our findings, and summarize our main results in Section~\ref{sec:conclusion_a2}.

\begin{table*}
\caption{Overview of previous detections on WASP-127\,b
\label{tab:param_detects_a2}}
\begin{threeparttable}
\begin{tabular}{lccccccccc}
     \hline
     Studies & Instruments  &  \water & Na & K & Hazes & Grey Clouds & Others\\
     \hline
\cite{Palle2017}  & NOT/ALFOSC  &  $\cdots$  &  Y (hints)  &  N  &  Rayleigh slope  &  MC/PC\expref{a} &  Hints of TiO, VO \\ 
\cite{Chen2018}  & GTC/OSIRIS  &  Y (hints)  &  Y  &  Y  &  Y  &   MC/PC\expref{a} &  Li \\ 
\cite{Zak2019}  & ESO/HARPS  &  $\cdots$  &  Y  &  $\cdots$  &  $\cdots$  &  $\cdots$  &  $\cdots$ \\ 
\cite{Welbanks2019}  & GTC/OSIRIS  &  Y (weak)  &  Y  &  Y  & $\cdots$   & $\cdots$   & $\cdots$  \\ 
\cite{Seidel2020_wasp127}  & ESO/HARPS  &  $\cdots$  &  N  &  $\cdots$  &  $\cdots$  &  $\cdots$  &  $\cdots$ \\ 
\cite{Allart2020}  & VLT/ESPRESSO  &  N  &  Y  &  N  &  $\cdots$  & Y    &  No Li or H \\ 
\cite{Santos2020}  & GST/Phoenix  &  $\cdots$  &  $\cdots$  &  $\cdots$  &  $\cdots$  &  $\cdots$  &  No He \\ 
\cite{Skaf2020}  & HST/WFC3  &  Y  &  $\cdots$  &  $\cdots$  &  $\cdots$  &  Y  &  FeH  \\ 
\cite{Spake2021}  & HST+\emph{Spitzer}  &  Y  &  Y  &  N  &  Y  &  N  &  \diox, but no Li  \\ 
\hline
\end{tabular}
\begin{tablenotes}
\item \textsc{\textbf{Notes}} --- Y for detected
, N for not detected
; \expref{a} MC/PC : Mostly clear -- Partly cloudy.
\end{tablenotes}
\end{threeparttable}
\end{table*}



\section{The unusual WASP-127 system}
\label{sec:wasp127}

\waspb\ \citep{Lam2017} is a highly-inflated sub-Saturn ($0.165 \pm 0.021\,M_{\rm Jup}$; \citealt{Seidel2020_wasp127}) that orbits a bright solar-type star (G5V, $H = 8.738$\,mag) with an orbital period of 4.17\,days (see Table~\ref{tab:param_sys_a2} for the stellar and planetary parameters of this system). 
Its radius ($1.31 \pm 0.03\,R_{\rm Jup}$; \citealt{Seidel2020_wasp127}) is anomalously large for a planet in such an old system ($9.7 \pm 1.0$\,Gyr; \citealt{Allart2020}) and places it on the edge of the short-period Neptune desert \citep{Mazeh2016}. The position of the star in a color-magnitude diagram indicates that it is moving off the main-sequence, and onto the RGB \citep{Lam2017}, 
and the associated increase in irradiation of the planet is thought to be the principal cause of the re-inflation of the planet's radius \citep{Lopez2016,Hartman2016}. 

However, other mechanisms have been proposed to explain the puffiness of short-period giant planets, such as tidal heating \citep{Bodenheimer2001,Bodenheimer2003}, enhanced atmospheric opacity \citep{Burrows2007}, or Ohmic heating \citep{Batygin2011, Thorngren2018}. 
Migration processes could have shaped the orbit through planet-planet interactions \citep[e.g.][]{Ford2008} or through the Kozai-Lidov effect \citep{Kozai1962, Lidov1962, Fabrycky2007, Bourrier2018}. 
The latter could also explain the retrograde misaligned orbit of \waspb\ ($\lambda = -128^{+6~\circ}_{-5}$; \citealt{Allart2020}), which further hints at a unusual formation and evolution pathway for this planet. Photo-evaporation was also suggested to partly explain the highly inflated state of \waspb\ \citep[][]{Owen2018}, but this is probably not the case due to the relatively low XUV flux that the planet receives from its old host star \citep{Chen2018, Santos2020}.

With an equilibrium temperature of 1400\,K, an extremely low density ($\sim 0.09$\,g\,cm$^{-3}$) and a low mean molecular weight around $2.3$\,g\,mol$^{-1}$ \citep{Skaf2020}, \waspb\ is very puffy, with an atmospheric scale height of about 2100\,km, making it an extremely favourable target for transit spectroscopy (the expected signal in transmission for one scale height is around 420 ppm; \citealt{Allart2020}). 
Contrary to other highly inflated exoplanets that show unexpectedly flat spectra \citep{Libby-Roberts2020}, previous atmospheric studies of \waspb\ have yielded strong detections of atomic and molecular species, with ground-based observations in the visible at both low- and high-resolution being fruitful, and are summarized in Table~\ref{tab:param_detects_a2}. The overall consensus seem to agree with WASP-127\,b having a partly cloudy atmosphere, with scattering hazes, water, and sodium.
%

The high-resolution ESPRESSO data by \cite{Allart2020} constrained 
the presence of a gray absorbing cloud deck to be between 0.3 and 0.5\,mbar (roughly consistent with the values from \citealt{Skaf2020} at $\log_{10}$ \cloud $= 1.85\inc{0.97}{0.66}\,$Pa $\simeq -3.15\,$bar). 
These same observations also enabled the measurement of the weak signature of the Rossiter-McLaughlin effect (RME; \citealt{Rossiter1924, McLaughlin1924}), the $v\sin(i_*)$ of the slowly rotating star, and the spin-orbit angle of the misaligned and retrograde planet. 

The richness of spectral information in \waspb's atmosphere has been unambiguously confirmed by \citet[][hereafter S21]{Spake2021}. They presented a combination of low-resolution spectroscopic transit observations with HST/WFC3 and the Space Telescope Imaging Spectrograph (STIS), as well as photometric observations with the Spitzer Space Telescope at 3.6 and 4.5\,\um. A retrieval analysis performed on their combined data revealed absorption from Na, \water, and \diox --- enabling the first firm constraints on \waspb's atmospheric metallicity and C/O ratio. Their analysis also revealed evidence for wavelength-dependent scattering from small-particle aerosols, however they did not detect K or Li, and found no evidence for a gray cloud deck.  
They tested both free\footnote{All individual molecular abundances are free parameters.}- and equilibrium-chemistry retrievals (with the frameworks ATMO, \citealt{Tremblin2015_ATMO, Tremblin2016_ATMO}, and NEMESIS, \citealt{Irwin2008_NEMESIS}) with both favoring super-solar abundances of \diox\ (compared to the expected value from a solar composition in chemical equilibirum, from an unusually strong absorption feature at 4.5\,\um; see their Figure~18), indicative of a high metallicity \citep{lodders_atmospheric_2002, oberg_excess_2016}. 
However, with only one photometric data point at 4.5\,\um, \citetalias{Spake2021} were unable to disentangle the contributions of CO and \diox~in a model-independent way. Due to this degeneracy, they found conflicting C/O ratios depending on the retrieval type. On the one hand, in the chemical equilibrium retrieval they found a super-solar C/O (roughly between 0.8 and 0.9), with an expected abundance of $\log_{10}\mathrm{CO} \simeq -2$ (see their Figure~12). On the other hand, in their free-chemistry retrieval, they found a sub-solar C/O ratio (below 0.2) as the abundance of CO remains unconstrained by the data. 
They concluded that spectroscopy with JWST will allow for precise constraints of on the C/O ratio or \waspb. 
Overall, a deeper investigation of this planet is needed to help shed light on how highly inflated planets form and evolve with time.

The previous strong detections of spectral features make \waspb\ a prime target for ground-based studies at high-resolution in the NIR. With no such studies having as of yet been published (although with an expected signal of 800--1000\,ppm of water in the NIR at high-resolution; \citealt{Allart2020}), we aim to bridge this gap with SPIRou \citep{Donati2020} and lift the degeneracy between the two previously proposed scenarios by detecting and quantifying the abundances of the main constituents of its atmosphere with transmission spectroscopy.

\begin{table}
\caption{WASP-127 System Parameters \label{tab:param_sys_a2}}
\begin{threeparttable}
\begin{tabular}{lc}
\hline
Stellar Parameters &  Value \\
\hline
Spectral Type\expref{a} & G5 \\
$H$ magnitude\expref{b} & 8.738\,mag \\
Stellar mass ($M_{*}$)\expref{c} & $0.950 \pm 0.023 \,M_{\odot}$ \\ 
Stellar radius ($R_{*}$)\expref{c} & $1.333 \pm 0.037\,R_{\odot}$ \\ 
Temperature ($T_{\rm eff}$)\expref{d} & $5842 \pm 14 \,$K \\
Surface gravity ($\log g$)\expref{d} & $4.23 \pm 0.02 \,$cgs \\
Metallicity ([Fe/H])\expref{d} & $-0.19 \pm 0.01$ \\ 
RV Semi-amplitude ($K_*$)\expref{c} &  $21.51 \pm 2.78$\,m\,s$^{-1}$ \\ 
$v \sin(i_*)$\expref{d} &  $0.53\inc{0.07}{0.05}$\,\kms \\
Age\expref{d} &  $9.7 \pm 1.0$\,Gyr \\
\hline
Planet Parameters &  Value\\
\hline
Planet mass ($M_{\rm P}$)\expref{c} & $0.165 \pm 0.021\,M_{\rm J}$ \\ 
Planet radius ($R_{\rm P}$)\expref{c} &  $1.311 \pm 0.025\,R_{\rm J}$ \\ 
Planet gravity ($g_{\rm P}$)\expref{d} & $236 \pm 32\,$cm\,s$^{-2}$ \\
Planet RV semi-amplitude ($K_{\rm P,0}$)\expref{d} & $129.95 \pm 1.2$\,\kms \\
Equilibirum temperature ($T_{\rm eq}$)\expref{a} & 1400\,K \\
Orbital period ($P$)\expref{c} & $4.178062(03)$\,days \\
Epoch of transit ( $T_0$)\expref{c} &  $2456776.62124(23)$ BJD \\ 
Transit duration ($T_{14}$)\expref{c} & $0.181(37)$\,days \\
Semi-major axis ($a$)\expref{c} & $0.04840\inc{0.00136}{0.00095}$\,AU \\
Inclination ($i_P$)\expref{e} & $87.85 \pm 0.35^{\circ}$ \\
Eccentricity ($e$)\expref{f} & $0$ \\ 
Systemic velocity Tr1 ($v_{\rm sys, 1}$)\expref{g} &  $-8.92 \pm 0.06$\,\kms \\ 
Systemic velocity Tr2 ($v_{\rm sys, 2}$)\expref{g} &  $-8.92 \pm 0.06$\,\kms \\ 
Systemic velocity Tr3 ($v_{\rm sys, 3}$)\expref{g} &  $-8.97 \pm 0.09$\,\kms \\   
\hline
\end{tabular}
\begin{tablenotes}
\item \textsc{\textbf{References}} --- \expref{a}\cite{Lam2017}; \expref{b}\cite{2MASS}; \expref{c}\cite{Seidel2020_wasp127}; \expref{d} $g_{\rm P}$ and $K_{\rm P,0}$ were computed by sampling 10,000 values of $R_{\rm P}$, $M_{\rm P}$, $M_{\rm *}$, $P$, and $i_{\rm P}$ from a normal distribution using their mean value and uncertainties listed here.; \expref{e}\cite{Allart2020}; \expref{f} Fixed;  \expref{g} This work.
\end{tablenotes}
\end{threeparttable}
\end{table}


\section{Observations}
\label{sec:Obs_a2}

In this work, we used the {\em Spectro-Polarim\`etre InfraRouge} (SPIRou; \citealt{Donati2020}), a fiber-fed \'echelle spectro-polarimeter installed at the Canada-France-Hawaii Telescope (CFHT, 3.6-m). SPIRou has a spectral resolution of $R = 70,000$ (with 2 pixels per resolution element, yielding a sampling precision of $\sim 2.3\,$\kms\ per pixel) and a broad NIR spectral range covering the $Y$, $J$, $H$ and $K_s$ bands simultaneously ($\sim$0.95--2.50\,\um). 
SPIRou splits the incoming target light into two orthogonal polarization states, each feeding its own science fiber (fibers A and B), allowing both states to be observed simultaneously. A third fiber (C) is fed by a calibration source.
The large NIR spectral range of SPIRou gives us access to the absorption features of a multitude of the major molecular species present in the atmospheres of exoplanets such as \water, \methane, \diox, HCN, and NH$_3$, but most importantly the 2.3\,\um\ CO band needed to differentiate the two scenarios presented in \citetalias{Spake2021}. The high-resolving power of SPIRou allows us to resolve and distinguish 
unique spectral line forests from different molecules (e.g., CO versus \diox), even if their bands overlap --- and its suitability for characterizing exoplanet atmospheres has already been demonstrated by several studies \citep[e.g.,][]{Pelletier2021, Boucher2021}. Of all existing facilities in the world to date, SPIRou, GIANO \citep{GIANO}, CRIRES+ \citep{CRIRES_plus}, IGRINS \citep{Park2014_IGRINS}, iSHELL \citep[][]{Rayner2012_ishell}, NIRSPEC \citep[][]{NIRSPEC}, and ARIES \citep{ McCarthy1998, Sarlot1999} cover the $K$-band and hence can lift the CO/\diox\ degeneracy which has hampered previous data. 

\begin{figure}
\hspace*{-0.2cm} 
\includegraphics[scale=0.43]{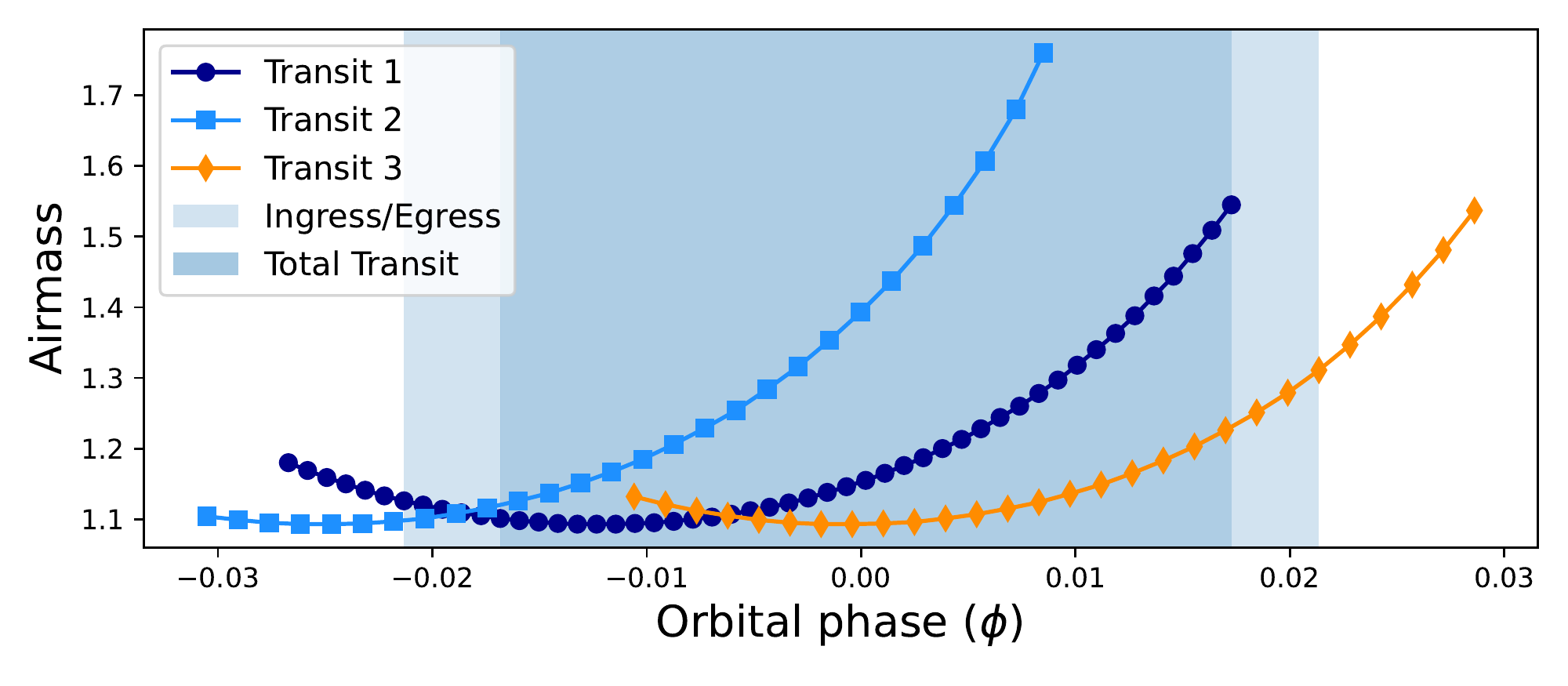}
\caption{\label{fig:AM_a2}
Airmass variation during the three SPIRou transit observations (dark blue circles for Transit 1, light blue squares for Transit 2 and orange diamonds for Transit 3. The shaded area shows the span of the transit event).  All three transits were taken under relatively favorable airmass conditions ($<$1.8).
} 
\end{figure}

\begin{figure}
\includegraphics[scale=0.41]{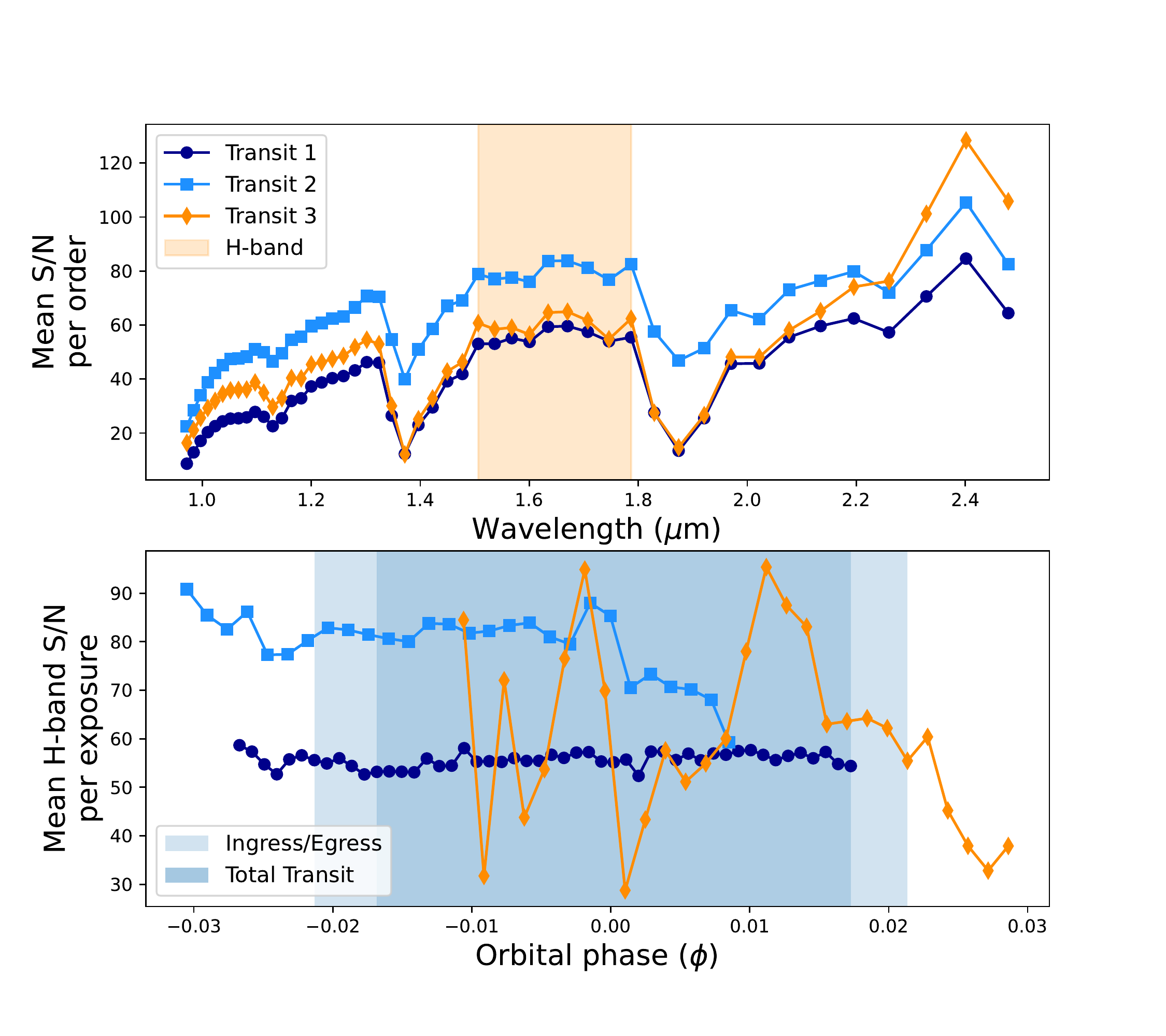}
\caption{\label{fig:S/N}
\emph{Top panel}: Temporal mean of the S/N per order for each of our three SPIRou transits. The shaded area highlights the orders in the $H$-band. \emph{Bottom panel}: Time evolution of the spectral mean of the S/N in the $H$-band per exposure, with the shaded area denoting the span of the transit event.  Here we see that while the first two transits were taken under relatively stable conditions, the Transit 3 data shows a much higher variability due to clouds/fog.
} 
\end{figure}

\begin{table*}
\caption{SPIRou observations of WASP-127 \label{tab:param_obs_a2}}
\begin{threeparttable}
\begin{tabular}{lccc}
\hline
Transit & Tr1 & Tr2 & Tr3 \\
\hline
UT Date   & 2020-03-11   & 2021-03-22 & 2021-05-03 \\
BJD (d)\expref{a}  & 2458919.85   & 2459295.86 & 2459337.72 \\ 
Texp (s)\expref{b} & 300 & 500 & 500  \\
Seeing ('')\expref{c} &  0.61--0.76 &  0.70--1.21 &  0.52--1.74\\
S/N\expref{d}  &   57.5 &  81.0 &  61.7 \\
\textbf{Number of exposures :} & & \\
Before ingress    &   7   &  7   &  0 \\
During transit    &  43   &  21  & 23 \\%
After egress      &   0   &  0   & 5 \\
Total             &  50   &  28  & 28 \\
Total observing time (h)\expref{e}      &  4.18  &  3.86 & 3.90\\
\hline
\end{tabular}
\begin{tablenotes}
\item \textsc{\textbf{Notes}} --- \expref{a} Barycentric Julian date at the start of the observing sequence
; \expref{b} Exposure time of a single exposure; \expref{c} Range of values of the seeing during the transit; \expref{d} Mean S/N per pixel, per exposure, at 1.7\,$\mu$m; \expref{e} Total observing time in hours.
\end{tablenotes}
\end{threeparttable}
\end{table*}

Three partial transits of \waspb\ were observed with SPIRou for a total of 11.94\,hours of observing time. All sets of observations were taken without moving the polarimeter retarders (Fresnel rhombs) to ensure the highest possible instrument stability, and placing the Fabry-P\'{e}rot in the calibration channel to better track small relative spectral drifts.

The observations of the first transit (hereafter Tr1) were obtained on 2020 March 11, as part of the SPIRou Legacy Survey (SLS). Initiated at the start of SPIRou operations, the SLS is a CFHT Large Program of 310 telescope nights (PI: Jean-François Donati; \citealt{Donati2020}) whose main goals are to search for planets around M dwarfs using precision RV measurements, characterize the magnetic fields of young low-mass stars and their impact on star and planet formation, and probe the atmosphere of exoplanets using HDS. 
The second and third transits (hereafter Tr2 and Tr3, respectively) were observed on 2021 March 22 and May 3 (Program 21AC02/21AF18, PI Boucher/Debras). 
Tr1 was observed with an exposure time of 300\,s, but this was increased to 500\,s for the following transits due to a lower than expected signal-to-noise ratio (S/N) seen in Tr1. The instrument rhombs were changed in August 2020 (between Tr1 and Tr2/Tr3), increasing the throughput in spectral bands $Y$ and $J$ by factors of 1.5 and 1.4, respectively, thus also improving the S/N in these bands for Tr2 and Tr3. 
An overview of the observation specifications are listed in Table~\ref{tab:param_obs_a2}. 
For each transit, the airmass curve is shown in Figure~\ref{fig:AM_a2}, while the S/N temporal mean per order and the spectral mean (over the $H$-band only) per exposure are shown in Figure~\ref{fig:S/N}. 
Sky conditions were photometric and dry for the first two observing sequences. However, fog caused Tr2's last exposure to be only 373\,s (instead of 500\,s). Nevertheless, we retain it for our analysis. For the third transit, poor weather conditions (higher water column density and mean extinction that varied between 0.2 and up to 1.8 mag, with an increase toward the end of the sequence) led to highly variable data quality and S/N. Nonetheless, the data reduction pipeline and estimation of uncertainties are reliable so we use the data anyway.


\section{Data Reduction and Analysis}
\label{sec:analysis_a2}

\begin{figure*}
\centering
\hspace*{-1cm} 
\includegraphics[width=1.1\textwidth]{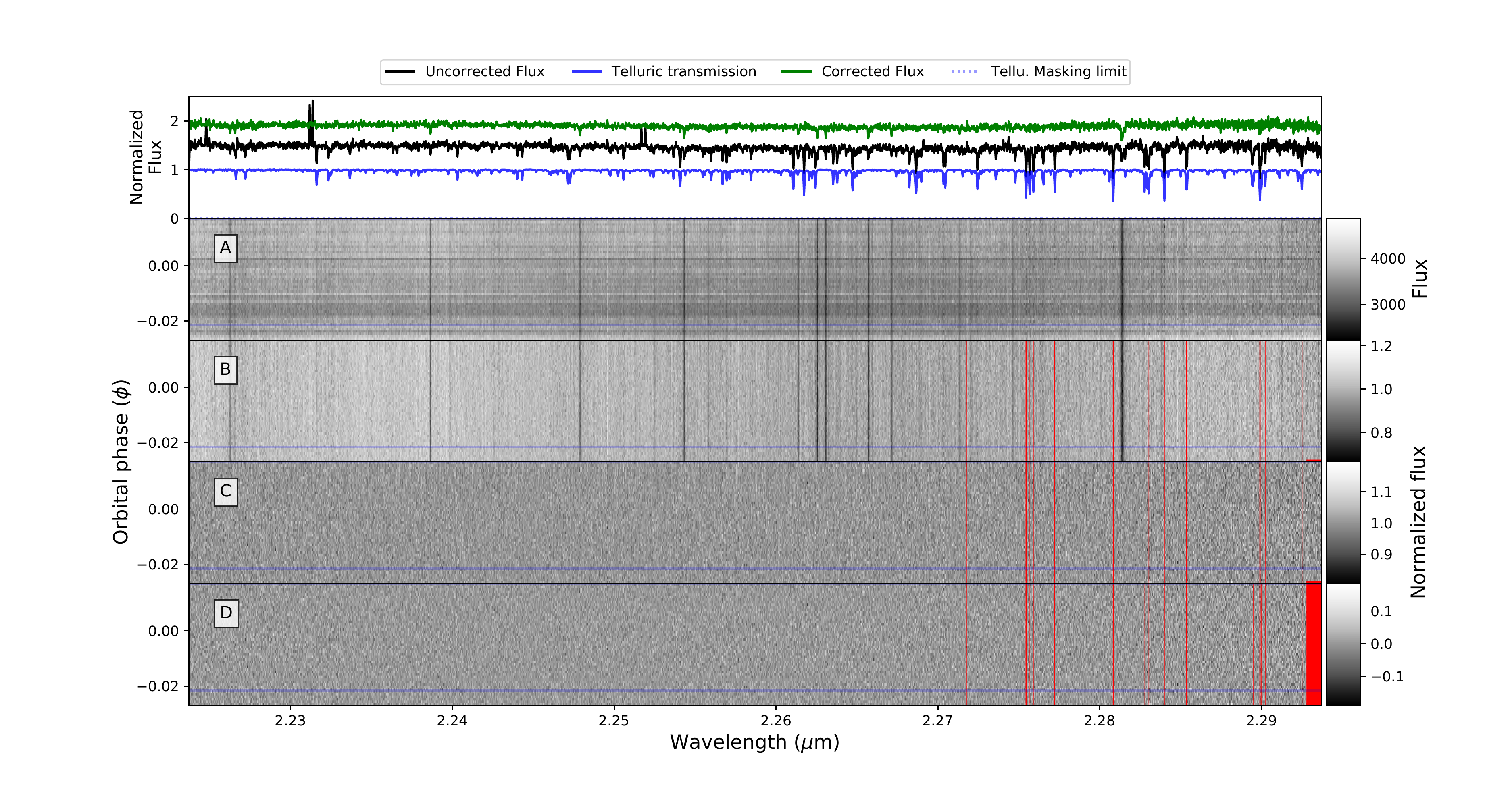}
\vspace{-0.7cm}
\caption{\label{fig:steps_a2}
Analysis steps that are applied to observed SPIRou spectral time series observations. Here the full order covering 2.2237--2.2937\,\um\ of Tr1 is shown. 
The top panel shows the uncorrected (black), the telluric-corrected (green) spectra, with offsets to facilitate visibility, and the reconstructed telluric transmission spectrum (blue). 
Panel (A): Telluric-corrected spectra. The masked pixels are shown in red; here the masking is done by \texttt{APERO}.
Panel (B): Spectra normalized to the continuum level of the reference spectrum. Most of the bad pixels were corrected or masked and the spectra were shifted in the pseudo star rest frame.
Panel (C): Transmission spectra, where each spectrum was divided by the reference spectrum.
Panel (D): Final transmission spectra, corrected for the vertical residual structures using PCA (2 principal components removed for Tr1). The total standard deviation of the transmission spectra time series is reduced by 3, 7 and 39\%, on average per order, from step C to D, for Tr1, Tr2, and Tr3, with 2, 2, and 5 PCs removed, respectively.
The blue horizontal lines in panels (A)--(D) show the ingress position (mid-transit is at phase 0).
} 
\end{figure*}

All data were reduced using {\em A PipelinE to Reduce Observations} (\texttt{APERO}; version 0.7.254; \citealt{Cook2022_APERO}), the SPIRou data reduction software. 
\texttt{APERO} performs all calibrations and pre-processing to remove dark, bad pixel, background and detector non-linearity corrections \citep{Artigau2018}, localization of the orders, geometric changes in the image plane, correction of the flat and blaze, hot pixel and cosmic ray correction, wavelength calibration (using both a hollow-cathode UNe lamp and the Fabry-Pérot etalon; \citealt{Hobson2021}), and removal of diffuse light from the reference fiber leaking into the science channels (when a Fabry-Pérot is used simultaneously in the calibration fiber). This is done using a combination of nightly calibrations and master calibrations. The result is an optimally extracted spectrum \citep{Horne1986} of dimensions 4088 pixels (4096 minus 8 reference pixels) with 49 orders (one order, \#80, is not extracted by the data reduction software), referred to as extracted 2D spectra; \textit{E2DS}. While the \textit{E2DS} are produced for the two science fibers individually (A and B) and for their combined flux (AB), we only used the AB extraction as this is the relevant data product for non-polarimetric observations.

\subsection{Telluric absorption correction}
\label{subsec:tell_rmv_a2}

\texttt{APERO} also provides telluric-corrected versions of the spectra, as well as the reconstruction of the Earth's transmittance (Figure~\ref{fig:steps_a2}, top panel). For our analysis, we use this data product to minimize contamination from telluric absorption (and sky emission). This step is essential to be able to recover the subtle signal of a exoplanet's atmosphere, especially as telluric contamination arises from molecules also expected to be present in exoplanet atmospheres.
The low number of observable transits of the \wasp's system (due to its low declination) prevented us from selecting optimal transits where the RV trail from the planet does not overlap with the telluric rest frame. Our observations thus have the potential to be contaminated by residual tellurics and their consideration and proper correction are even more crucial.  

The telluric correction done in \texttt{APERO} is a two-step process and is briefly outlined here (this also includes the sky emission correction). 
1) The extracted spectra of both science targets and a large set of rapidly rotating hot stars are fitted with an Earth’s transmittance model from TAPAS \citep{TAPAS} that leaves percent-level residuals. 2) From the ensemble of hot star observation, we derive a correction model for the residuals with three components for each pixel (optical depths for the H$_2$O, the dry components, and a constant). This residual model is adjusted to each science observation according to the optical depth of each component from the TAPAS fit. The resulting correction leaves residuals at the level of the PCA-based method of \cite{Artigau2014}, but has the advantage of a simplicity and that any spurious point in the data will result in a local error rather than affecting the transmission globally as for a PCA analysis. 
Finally, a reconstruction of the telluric spectrum is derived using the fitted TAPAS template and the residuals model, for each observed spectrum. 
%
The pipeline performs the telluric absorption correction for lines with a transmission down to $\sim10\%$ (i.e., with relative depths of 90\% with respect to the continuum), with deeper lines being masked out, and additional masking being at the discretion of the user. Following an injection-recovery test, we observed that some of the data yielded slightly better detections with further masking. We thus additionally masked telluric lines (from the reconstructed telluric spectra) with core transmission below 30\% to 55\% depending on the depth of tellurics for a given night, and this mask was extended from the core of those lines until their transmission in the wings reached 97\%. No additional masking on Tr1 was found to be necessary from injection-recovery tests.


\subsection{Transmission spectrum construction}
\label{subsec:transpec_a2}

To build the transmission spectra, we follow a similar process to \cite{Boucher2021}, and apply the technique individually to each order (and each transit). We briefly summarize the process here.
\begin{itemize}
    \item Bad pixels correction and masking: The telluric-corrected \textit{E2DS} spectra are blaze normalized (Figure~\ref{fig:steps_a2}\,A), then the bad pixels are identified and corrected based on the method of \cite{brogi_carbon_2014}, where the spectra are divided by their mean value (wavelength-wise, to bring everything to the same continuum level) and then each spectral pixel is divided by its mean in time, yielding a data residual map, i.e., a noise map. We take the absolute value of the map and fit a second order polynomial to each exposure. This recovers the noise floor, which follows the noisier borders of each order, and we subtract it from our noise map. Then, any pixels deviating by more than 5$\sigma$ from the flat noise map are flagged as bad pixels. For isolated pixels, they are corrected with a spline-interpolation, while groups of 2--3 pixels are linearly interpolated. Groups of four or more pixels are masked (less than 0.001\% of the pixels are masked in this step for all transits). We follow by masking spectral pixels still deviating by more than 4$\sigma$ in the time direction within each order\footnote{Here, the removal of the noise floor, to account for the natural increase of noise near the order borders, was also applied.}, and mask the neighboring pixels until the noise level reaches 3$\sigma$ (roughly 1\% of the pixels are masked in this step for each transit). 
    
    \item Stellar signal alignment: After the bad pixel corrections, we return to the wavelength-normalized spectra and Doppler shift each spectrum to a pseudo stellar rest frame (SRF), where the stellar lines are aligned, but not centered at zero velocity (Figure~\ref{fig:steps_a2}\,B). We accomplish this by shifting only by the negative of the \emph{variation} of the barycentric Earth RV and stellar orbital motion\footnote{Stellar orbital motion was included for completeness, even if it is negligible.}. This RV variation from each exposure is compared to the value from the middle of each transit sequence. This means that the first half of the exposures are shifted by roughly the same amount as the second half, but in the opposite direction, therefore minimizing interpolation errors caused by shifts of large fractions of pixels. In this case, the shifts are at most 12\% of a spectral pixel on the SPIRou detector. We then mask the telluric lines as described above.

    \item Reference spectrum construction and removal: Next, we build a reference spectrum representative of the stellar spectrum to remove its contribution from each observation and leave only the planetary signal. The usual procedure is to only use the out-of-transit exposures to construct this reference spectrum, thus minimizing any contamination by the planetary signal. However, given that our observations only include a small number of low-S/N exposures, and only cover either ingress or egress, a reference spectrum constructed in this way would not be a true representation of our entire observing sequence. 
    For this reason, we construct the reference spectrum by taking the median of \emph{all} spectra (in the pseudo SRF). Additionally, for Tr3, due to the low S/N of certain exposures, we build the reference spectrum using only those exposures with mean $H$ band S/N $\geq 40$.
    These final reference spectra should not contain much residual planet signal as the latter move over roughly 4.3, 3.3 and 3.6 times the line full width at half maximum (FWHM) during the transit for Tr1, Tr2 and Tr3 respectively, compared to the quasi-stationary stellar signal. Dividing by the reference spectra leaves low-frequency variations, that can be corrected by dividing the spectra by a low-pass filtered\footnote{Median filter of width 51 pixels followed by convolution with a Gaussian kernel of width 5 pixels.} version of the spectra divided by the reference. 
    
    \item We observed in some parts of some spectra that this ratio was far from 1.0, meaning that these parts are more poorly represented by the reference spectrum. We thus chose to mask the regions where this ratio was 6\,$\sigma$ away from the mean value, which removes less than 1\% of the data points for all transits. 
    
    \item The individual transmission spectra are obtained by dividing the continuum-normalized spectra by the reference spectrum (Figure~\ref{fig:steps_a2}\,C). 
    We then reapply a final sigma-clipping at 6\,$\sigma$ to insure the removal of any remaining outlier pixels before the next step (masking less than 0.5\% of the data). 
    
    \item Systematic noise residuals correction: We used a PCA-based approach to remove any remaining pseudo-static signals (e.g., stellar and telluric residuals). We build the PC base in the time direction using the natural logarithm of the transmission spectra themselves, and then perform injection/recovery tests (at the negative $K_{\rm P,0}$) to determine the appropriate number of PCs to remove. A combination of the best retrieved $t$-test, CCF SNR and/or $\Delta$~BIC (see subsection~\ref{subsec:detec-signif}) are considered to make the selection, as these metrics can all suggest slightly different optimal PCA prescriptions. This results in the removal of \pcun, \pcdeux, and \pctrois\ PCs for Tr1, Tr2, and Tr3, respectively. The optimal number of necessary PCs seems to follow the data quality and S/N (i.e., Tr3 needs more PCs to better uncover an injected underlying signal). 
    We also tested performing our analysis using the telluric-\textit{uncorrected} spectra, and only using a PCA-based approach to remove unwanted stationary (or quasi-stationary) contributions. However, this did not perform as well at removing telluric residuals as the two-step approach that we adopted here. 
    
    \item 
    Finally, we remove the mean of each spectrum (wavelength-wise) to keep a zero mean for the computation of the cross-correlation function (see Fig.~\ref{fig:steps_a2}\,D). 
\end{itemize}

We measured \vsys\ directly from our data by computing the CCF of the telluric-corrected 1D spectra ($S1D$; another data product from \texttt{APERO}) with a synthetic spectrum from a PHOENIX atmospheric model \citep{Husser2013} with $T_{\rm eff} = 5800\,$K, $\log g = 4.0$, and $[M/H]=0$. We computed the CCF with the signal in $H$ band of every $S1D$, weighted by the second derivative of the model (a proxy for the strength of the absorption lines), and measured the peak position with the bisector method. We then subtracted $v_{\rm bary}(t)$ and $v_{\rm orb,*}(t)$, (where $v_{\rm bary}(t)$ is the barycentric velocity of the observer --- in our case it is the barycentric Earth RV, BERV --- and $v_{\rm orb,*}(t) = - K_* \sin [2\pi(\phi(t))]$ is the radial part of the orbital velocity of the star), and took the mean over all spectra of each transit to get the observed \vsys\ (the values are listed in Table~\ref{tab:param_sys_a2}). 
We computed the gravitational redshift to be at $0.47\inc{0.01}{0.02}$\,\kms, and estimated the convective blue shift to be $-0.42 \pm 0.15$\,\kms \citep{Leao2019}. As these values roughly compensate one another, we chose to simply ignore these effects in the determination of the heliocentric RV of the \wasp\ system.


\section{Atmospheric Signal Extraction: Models and Methods} 
\label{sec:correl_a2}

Even though the data are now cleaned of telluric and stellar signals, the individual planetary absorption lines are still buried within the remaining noise. We thus need a cross-correlation type analysis that combines the signal of all lines over the available spectral range to reveal the planetary atmospheric signal. In this section, we first present how \waspb's atmospheric models are generated and how we process them to better represent the data. Then, we present the methods that we tested to detect the atmosphere of \waspb: cross-correlation, \textit{t}-test, and log-likelihood mapping.

\subsection{Atmospheric Model}
\label{subsec:model_a2}
We generated synthetic transmission spectra of \waspb's atmosphere using the open-source \texttt{petitRADTRANS} framework \citep[PRT; ][]{petitRADTRANS, molliere_retrieving_2020}. PRT computes transmission and emission spectra of exoplanets with clear or cloudy atmospheres. It can produce low-resolution ($R=1,000$) or high-resolution ($R=10^6$) models by considering the molecular opacities at each pressure layer using either a correlated-k treatment or line-by-line radiative transfer, respectively. In this work we used 50 pressure layers between $10^{-10}$ and $10^2$\,bar, log-uniformly spaced, and fixed the reference pressure to $P_0 = 10$\,mbar\footnote{This is where the planetary radius is equal to $R_{\rm P}$.}, where the bulk of our transmission signal is expected to originate. The molecular opacities and associated line lists used in this work include\footnote{A complete list of the available opacities is given in the documentation of PRT at: \url{https://petitradtrans.readthedocs.io/en/latest/index.html}. Also, we added some high-resolution opacities to PRT manually (OH and FeH) following their instruction, and used the open-access DACE database, computed with HELIOS-K \citep[][]{grimm_helios-k_2015, grimm_helios-k_2021}, to compute the opacity grids : \url{https://dace.unige.ch/opacityDatabase/}} \water, CO, \diox, and OH\ \citep{HITEMP2010}, \methane\ \citep{yurchenko_exomol_2020}, HCN \citep{barber_exomol_2014, harris_improved_2006}, NH$_3$ \citep{yurchenko_variationally_2011}, FeH 
\citep{Bernath2020_mollist}, TiO \citep[see references in ][]{petitRADTRANS}, and C$_2$H$_2$ \citep{Rothman2013_hitran}. Out of these elements, only \water, CO, \diox, \methane, NH$_3$, and TiO are exepected to have abundances larger than $10^{-10}$ for a solar composition in chemical equilibrium.
All these molecules have major features within SPIRou's spectral range that could potentially be detected given a high enough volume mixing ratio (VMR). Also included are the absorption from H$_2$ broadening \citep{Burrows2003}, collision-induced broadening from H$_2$/H$_2$ and H$_2$/He collisions \citep{Borysow2002}. The abundances of H$_2$ and He were fixed to 85 and 15\% (i.e., solar-like), respectively. To save time on the generation of the high-resolution PRT models, we down-sampled the line-by-line opacities by a factor of 4, which gave us models at $R = 250,000$.

We adopted an analytical atmospheric temperature-pressure (T-P) profile from \cite{Guillot2010}, who derived a parametrized relation between temperature and optical depth valid for plane-parallel static grey atmospheres. For simplicity and to anchor the shape of the profile to that of \citetalias{Spake2021}, we fixed three of the four parameters of the profile, namely $\kappa_{\rm IR}$ the atmospheric opacity in the IR wavelengths, $\gamma$ the ratio of the optical and IR opacities, and $T_{\rm int}$ the planetary internal temperature, while keeping $T_{\rm eq}$, the atmospheric equilibrium temperature, as a free parameter\footnote{We will interchange the references to $T_{\rm eq}$ and $T_{\rm P}$.}. We fixed the above parameters to $\kappa_{\rm IR} = 10^{-3}$\,cm\,g$^{-1}$, $\gamma = 10^{-1.5}$ and $T_{\rm int} = 500$\,K, based on the values retrieved by \citetalias{Spake2021} using ATMO, and do not change them for the entirety of this work. 

We include a gray cloud deck whose contribution is characterized by its cloud-top pressure \cloud. In order to model the chromatic absorption by aerosols ($\sigma_{\rm haze}$), we used a power law of the form : 
\begin{equation}\label{eq:scat}
    \sigma_{\rm haze}(\lambda) = S \sigma_0 \left( \frac{\lambda}{\lambda_0} \right) ^{\gamma},
\end{equation}
where $S$ is the empirical enhancement factor, $\sigma_0$ is the scattering cross-section of molecular hydrogen at $\lambda_0 = $ 0.35\,\um, and $\gamma$ sets the wavelength dependence of the scattering.

\subsubsection{Rotation kernel}
\label{subsubsec:rot_ker}

We explored the possibility of WASP-127\,b's atmospheric signal being broadened due to tidally-locked rotation. For this, we used a rotation kernel built from a parameterization of the rigid body rotation of a shell (here representing the atmosphere, valid in the transit geometry) as developed in Annex~\ref{an:rot_ker}. 
The shape of the kernel is dictated by the thickness of the shell (related to the scale height of the atmosphere, thus related to $M_{\rm P}$, $R_{\rm P}$, $T_{\rm P}$, and the atmosphere's mean molecular weight, $\mu$), and its rotation frequency ($\omega$). We also added the possibility to vary the relative kernel intensity of the two hemispheres by a certain fraction $f$\footnote{$f$ is defined as the fraction of the morning (leading) limb over the evening (trailing) limb. $f<1$ implies a muted contribution from the morning limb compared to the evening limb.}, which would allow for asymmetrical integrated speeds between the evening and morning side of the terminator. Physically, this asymmetry could originate from clouds masking the contribution of certain parts of the atmosphere \citep[e.g.,][]{Ehrenreich2020, Savel2022_wasp76}. 

Once generated, the model is then convolved with this kernel, that has been normalized beforehand, and binned to match SPIRou's sampling.

\subsubsection{Model processing}
\label{subsec:mod_seq_a2}

At this point in the analysis, all the detrending steps applied to the data have not only affected the telluric and stellar signals, but also to some extent, that of the planetary atmosphere \citep[][]{Brogi2019}. 
This is mostly due to the subtraction of the PCs, which are generally not fully orthogonal to the planet's transmission spectrum. Their subtraction (see Section~\ref{subsec:transpec_a2}) may warp part of the actual planet signal and introduce artifacts in the spectral time series which can then bias the determination of atmospheric parameters (velocities, abundances, temperature, etc.). We thus apply the same treatment to the model before comparing it to the data to ensure a better representation. 
This method was first presented in  \cite{Brogi2019}, was successfully used in many studies and has since become a standard procedure for this type of analysis  \citep[e.g.,][]{gandhi_hydra-h_2019,  Pelletier2021, Boucher2021, Kasper2021, Line2021, Gibson2022}. 
We proceed as follows to generate full synthetic transit sequences:  
we produce a model spectrum $1-(R_{\rm P}(\lambda)/R_*)^2$, which we then inject at \vvv{P}$(t)$, the total planet RV, 
\begin{equation}
v_{\rm P}(t) =  v_{\rm bary}(t) + v_{\rm sys} + v_{\rm orb, P}(t) - v_{\rm orb, *}(t) +  v_{\rm rad}\,,
\label{eq:v_planet_a2}
\end{equation}
where \vsys\ is the systemic RV; $v_{\rm orb, P}(t) = K_{\rm P} \sin [2\pi(\phi(t))]$ is the radial part of the orbital velocity of the planet; \kp\ and $K_*$ are the planet's and star's RV semi-amplitude, and $v_{\rm rad}$ is a constant additional velocity term to account for potential shifts. Note that here, \vvv{bary}$(t) =$\,\vvv{bary}$(t_{\rm mid. exp.})$, the barycentric velocity at the middle of the sequence, because the spectra are in the pseudo-SRF (from the second step in Section \ref{subsec:transpec_a2}).  
We can then generate synthetic time series for different combinations of \kp\ and \vvv{rad} to produce different planetary paths in velocity space. 
We then project the PCs on the synthetic transit sequence, i.e. compute the best fit coefficients from the PC base obtained with the real observations, and remove this projection from the model sequence. 
This procedure serves to replicate on our model, to the extent possible, any alteration of the real planetary signal that occurs during the data analysis \citep{Brogi2019}.
We note that the non-PCA reduction steps do not need to be re-applied to the synthetic sequence as they have no impact on the planetary transmission. After this PC subtraction (done in log-space), we remove the mean spectrum for each order, as was done on the observed data. This synthetic, PC- and mean-subtracted model transmission time series ($m_i$) is then used for the computation of the cross-correlation and log-likelihood. 


\subsection{Cross-Correlation and Log-Likelihood Mapping}
\label{subsec:CCF_a2}

To extract and maximize the planetary signal, we combine the signal of the many buried, but resolved absorption lines that are found over the whole spectral range of SPIRou with cross-correlation and similar approaches. 
The approach that we used to detect the planet signal and constrain the atmospheric parameters is described below.

\subsubsection{Algorithm} 

Based on the equations in \cite{Gibson2020} (also used in \citealt{Nugroho2020, Nugroho2021, Boucher2021, Gibson2022}), we write the cross-correlation function (CCF) as

\begin{equation}
\mathrm{CCF}(\boldsymbol{\theta}, v_{\rm P}) = \sum^N_{i=1} \frac{ f_i \cdot m_i(\boldsymbol{\theta}, v_{\rm P})}{\sigma_i^2}\,,  
\label{eq:CCF_a2}
\end{equation}
which is equivalent to a weighted CCF, where $f_i$ are the transmission spectra (described in Section~\ref{subsec:transpec_a2}) with associated uncertainties $\sigma_i$, $m_i$ is the model (described in Sections~\ref{subsec:model_a2} and \ref{subsec:mod_seq_a2}), and $\boldsymbol{\theta}$ is the model parameter vector, which includes the atmospheric model parameters, and generated at a given orbital solution $v_{\rm P}$. The index $i$ runs over all times and wavelengths in the data set, and the summation is done over $N$ data points (total number of unmasked pixels).

The uncertainties $\sigma_i$ were determined by first calculating, for each spectral pixel, the standard deviation over time of the $f_i$ values. 
This provides an empirical measure of the relative noise across the spectral pixels which captures not only the variance due to photon noise but also due to such effects as the telluric and background subtraction residuals, but it does not convey how the noise inherent to one spectrum compares to that of another. To capture this latter effect and include it in the $\sigma_i$, the dispersion values calculated above were multiplied, for each spectrum, by the ratio of the median relative photon noise of that spectrum divided by the median relative photon noise of all spectra. This is computed prior to normalization: the S/N variations across the night and the different orders are thus accounted for in this $\sigma_i$ term, acting as a weight. The final uncertainty values $\sigma_i$ thus reflect both temporal and spectral variability. 
We compared with another method to compute the uncertainties to validate ours. For this, we used a $\sigma$ map following the \cite{Gibson2020} method, i.e., by optimizing a Poisson noise function of the form $\sigma_i = \sqrt{aF_i + b}$ with the residuals from removing the first 5 PCs. Since it yielded very similar results, we continued with our method.

The CCF (equation~\ref{eq:CCF_a2}) is calculated for every order of every spectrum for an array of \vvv{rad} of size $n_v$. This gives a cube with size $106 \times 49 \times n_v$ (when combining all three transits\footnote{We combine the transits by simply concatenating the three CCF time series, order-per-order. There are 50 spectra in Tr1 and 28 in Tr2 and Tr3, for a total of 106.}) for a given \kp\ value in the modeled sequence and for each model tested. 
To combine everything into a single CCF, we first sum the above cube over orders, and then over time by applying a weight to each spectrum according to a transit model (transit depth at a given time), computed with the \texttt{BATMAN} package \citep{Kreidberg2015_batman}. We used a non-linear limb darkening law from \cite{Claret2000} with fixed coefficients $u_1 =0.5944$, $u_2 = 0.0707$, $u_3 = -0.1204$, and $u_4 = 0.0202$, taken in \citetalias{Spake2021}, and valid 
for their white light-curve fit using WFC3+G141. The ephemeris and system parameters used are listed in Table~\ref{tab:param_sys_a2}.


The CCF equation above can then be used to compute the S/N and/or the $t$-test (see next Section) and estimate the level of detection of a given signal. However, to get a better model selection method, we start from equation~\ref{eq:CCF_a2} and then map it to a $\beta$-optimised likelihood function, as presented in \cite{Gibson2020}\footnote{The additional scaling factor $\alpha$, which accounts for any scaling uncertainties of the model was set to 1 following \citep{Brogi2019}, because we do not expect much signal to come from an extended exosphere. Also, this parameter is correlated with the VMR, \tp, and \cloud, and including it just adds an additional degeneracy.}, which is a more generalized form of the likelihood function in \cite{Brogi2019}:
\begin{equation}
\ln \mathcal{L} = -\frac{N}{2} \ln\left[ \frac{1}{N} \left( 
\sum \frac{f_i^2}{\sigma_i^2} + \sum \frac{m_i^2}{\sigma_i^2}
- 2 \sum \frac{f_i m_i}{\sigma_i^2} \right) \right] 
\label{eq:loglikelihood_a2}
\end{equation}
where the summation is implied over $i$ (both spectral pixels and time). 
This equation can be written in a more compact form, i.e.:
\begin{equation}
\ln \mathcal{L} = -\frac{N}{2} \ln  \frac{\chi^2}{N},
\label{eq:logl2chi2_a2}
\end{equation}
by using the definition of the $\chi^2$:
\begin{equation}
\chi^2 = \sum \frac{(f_i-m_i)^2}{\sigma_i^2}.
\label{eq:chi2_a2}
\end{equation}

\subsubsection{Detection significance}
\label{subsec:detec-signif}

There exist multiple methods to quantify the detection significance of a signal --- here, we present the three that we computed: the S/N of the CCF peak, the $t$-test, and $\Delta$\,BIC.

First, the ``S/N significance'' is determined by dividing the total CCF (either the 2D \kp\ versus \vvv{rad} map, or the 1D version varying only with \vvv{rad}) by its standard deviation, the latter being calculated by excluding the region around the peak ($\pm 15$\,\kms\ in the \vvv{rad} space and $\pm 70$\,\kms\ in the \kp\ space). 
This is a useful metric to quickly assess whether or not a detection is significant, but its value should be viewed as only approximate as it can vary highly with the \vvv{rad}\ and \kp\ range and sampling, the extent of the excluded peak region, whether it is computed on the 2D map or the 1D CCF, etc. \citep[][]{Cabot2019}.

The second is the Welch \textit{t}-test \citep{Welch1947}, which has been used in many previous studies  \citep[e.g.,][]{Birkby2013, Birkby2017, Brogi2018, Cabot2019, Alonso2019_water, Webb2020, Boucher2021, Giacobbe2021}. The \textit{t}-test verifies the null hypothesis that two Gaussian distributions have the same mean value. The Welch $t$-test is a generalisation of the Student test \citep{Student1908}, for which the two samples can have unequal variances and/or unequal sample sizes. In our case, the two distributions to be compared are drawn from our correlation map. On the one hand, we have the in-trail distribution of CCF values, that is, the CCF values within 3-pixels wide columns centered on the peak (i.e., the signal following the planet RV path; as done in \citealt{Birkby2017} and suggested in \citealt{Cabot2019}). On the other hand, we have the out-of-trail distribution, which includes the CCF values more than 10\,\kms\ away from $v_{\rm P}(t)$, where there should be no planet signal. The \textit{t}-test then evaluates the likelihood that these two samples were drawn from the same distribution. This approach is more robust against outliers than the CCF SNR, and provides a complementary assessment of the detection significance. 
However, this metric might still not reflect the true confidence in a molecular detection in absolute terms since, as pointed out in \cite{Cabot2019}, it is not robust against oversampling in the velocity space, and thus somewhat arbitrary.

The third is the Bayesian Information Criterion (BIC)\footnote{BIC$ = k\ln(n) - 2 \ln\mathcal{L}$; where $k$ is the number of parameters, $n$ is the number of data and \logl\ is our log-likelihood value for our model for each combination of parameters. For fixed values of $k$ and $n$, the lowest BIC corresponds to the highest \logl.}, which we applied to our \logl\ results to establish how the best-fit model fares compared to the others. In this formalism, the model with the lowest BIC is preferred (here, taken to be the best-fit model), and the evidence against models with higher BIC is usually described as ``very strong" when $\Delta$\,BIC $= 2\,\Delta $\logl\ is greater than 10 \citep{Kass1995}. 

We note here that in all following calculations, we exclude the exposures where the planet velocity is within 2.3\,\kms\ (1 pixel) of the BERV in order to prevent any potential telluric contamination of the planet signal. This leads to the removal of the last six and three exposures of Tr1 and Tr2 respectively, and the first three of Tr3. 

\section{Results}
\label{sec:results_a2}

We first applied the CCF, $t$-test, and \logl\ analyses to our SPIRou \waspb\ transmission data. We used models consisting of H, He, and one of the individual molecules that are usually present in the atmospheres of giant planets, namely \water, CO, \diox, \methane, HCN, NH$_3$, and C$_2$H$_2$ \citep{lodders_atmospheric_2002}.
For completeness, we also included FeH, TiO, and OH. We cycled through the aforementioned molecules one at a time, and for each case, we tested VMRs between $10^{-9}$ and $10^{-2}$. We obtained a clear detection of \water, as well as a tentative detection of OH. 

\begin{figure}
\hspace*{-0.4cm}
\includegraphics[scale=0.43]{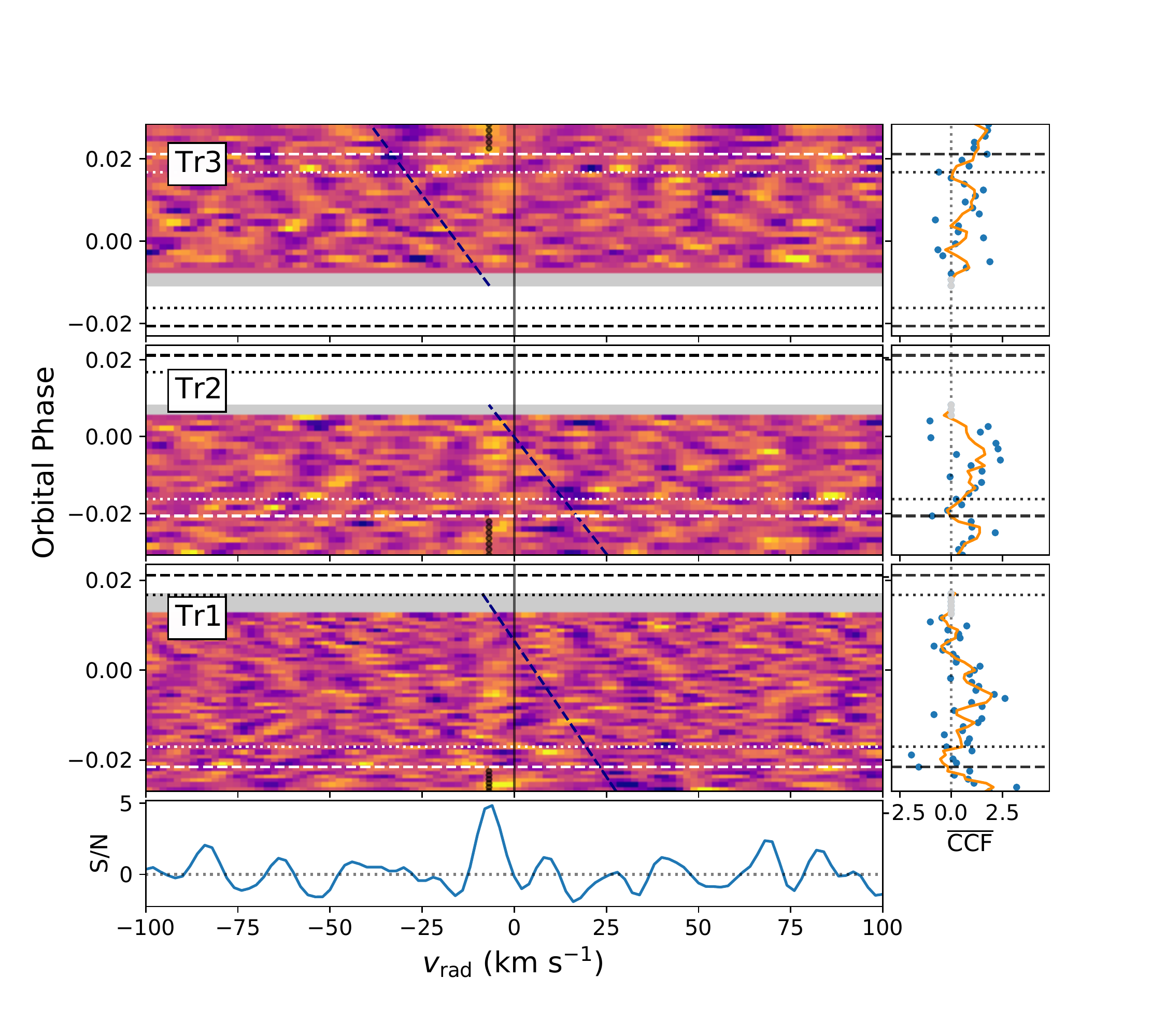}
\vspace{-0.5cm}
\caption{ Planet rest-frame cross-correlation time series of the best-fit \water\ model for the three SPIRou transits of \waspb.  
(Left panels) Normalized CCFs from individual exposures as a function of \vvv{rad} for Tr3 (first panel), Tr2 (second panel) and Tr1 (third panel), in the planetary rest frame (\kp $=K_{\rm P, 0}$). The diagonal navy dashed lines show the position of the BERV, while the horizontal 
dashed (and dotted) lines show the ingress and egress (and $T_{2,3}$) positions. 
The black vertical lines show the planet path for \vvv{rad} $ = 0$\,\kms\ (full line) and \vvv{rad} = \vvv{peak} (dotted line --- only shown out-of-transit to increase visibility during transit).
The exposures where the BERV crosses the peak position by less than 2.3\,\kms\ are excluded, and shown in grey. The bottom panel shows the 1D CCF S/N curve for the combined transits. 
(Right panels) Mean CCF for a 3-pixel column bin centered on \vvv{peak} (blue points) and the 3-exposures binned signal (orange line). The grey dots show the excluded exposures. }
\label{fig:CCF_2d_a2}
\end{figure}

\begin{figure}
\hspace*{-0.2cm}
\includegraphics[width=0.49\textwidth]{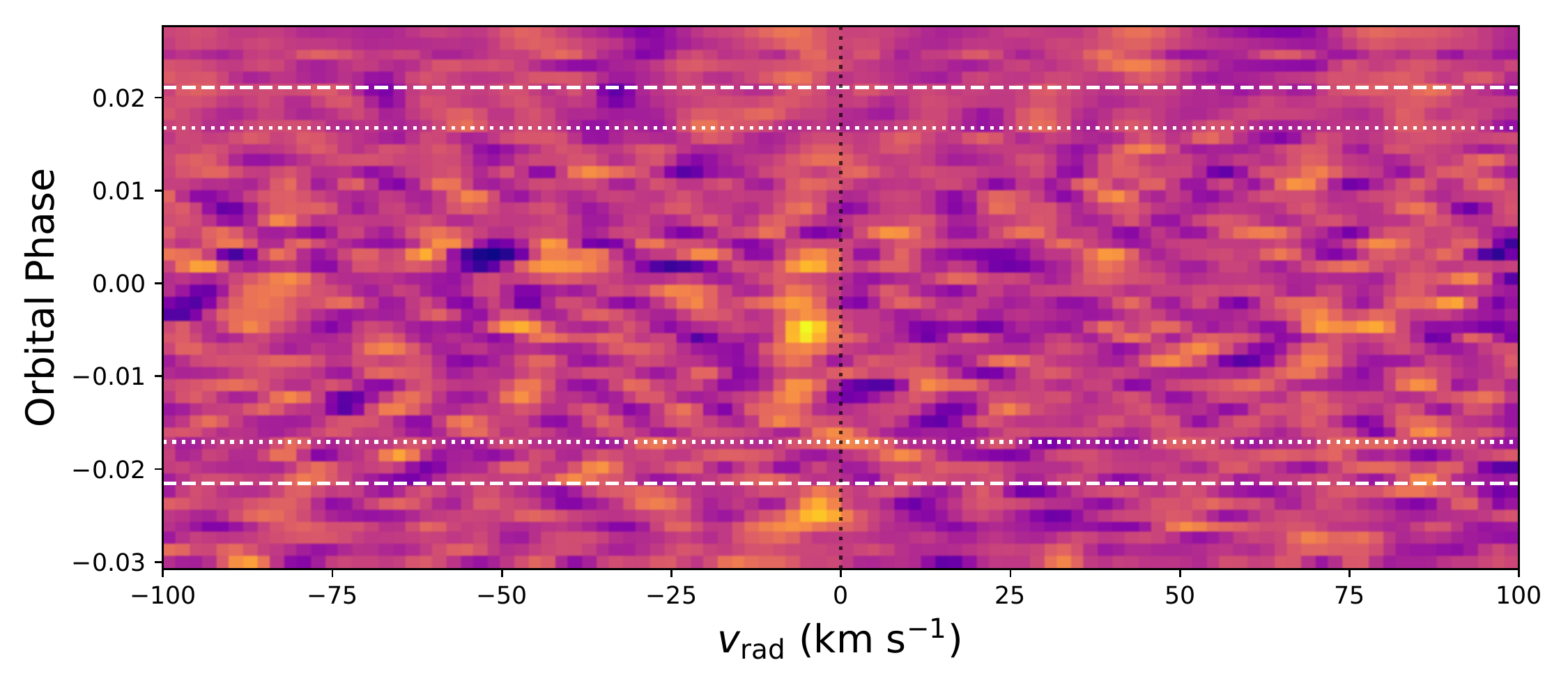}
\vspace{-0.5cm}
\caption{Same as Figure~\ref{fig:CCF_2d_a2}, but where the CCFs of the different transits were linearly interpolated on a common orbital phase grid, then summed.}
\label{fig:CCF_2d_sum}
\end{figure}

To further quantify the \water\ detection, we ran a series of tests on a grid of models across a broad range of \vmr, $T_{\rm eq}$, and \cloud\ to find the model which maximized the detection significance. 
According to the $t$-test, the best-fit model has $\log_{10}$ \water$ = -5$, $T_{\rm eq} = 1200$\,K and \cloud$ = 0.1$\,bar. The CCF time-series in the planet's rest frame, at \kp $ = K_{\rm P,0} = 129.95 \pm 1.2\,$\kms\ (the expected position of the planet), is shown in Figure~\ref{fig:CCF_2d_a2}, and a combined version is shown in Figure~\ref{fig:CCF_2d_sum}. 
On the CCF time-series (Figure~\ref{fig:CCF_2d_a2}), there are noise structures in the out-of-transit exposures at similar \vvv{rad} as the planetary signal that mimics an extended signal. This is most likely some residual noise from the data reduction that simply appears close to \vvv{peak}. Especially in Tr1, the first few exposures are much noisier than the rest and we can see similar (negative) amplitudes at other velocities. Plus, the structures are extended, and deviate from the vertical (planetary) path.
For Tr3, the last few exposures are affected by fog emergence, and a S/N drop, which ended the observations prematurely. Hence, these residuals can naturally be explained by the poor quality of Tr3. 
These noisy exposures always show residuals irrespective of the fraction of tellurics masked, number of PCs removed, or version of \texttt{APERO} used.
These exposures also appear noisier than the rest in the CCF map of OH and other non-detected molecules, such as \diox, which further suggests that some residuals are present out-of-transit. We found that removing a second-order polynomial in the time direction was able to remove these structures (while still keeping the in-transit planetary atmosphere signal). However, to stay consistent, we would need to apply this step to the generated model sequences, and recomputing the coefficients for every tested model would be too computationally expensive for retrievals. Therefore, we did not apply the polynomial removal step. Nonetheless, since the residuals are mostly outside of the transit, they are not used for the likelihood evaluation and thus should not affect the retrieval results.

\begin{figure}
\centering
\hspace*{-0.5cm}
\includegraphics[scale=0.47]{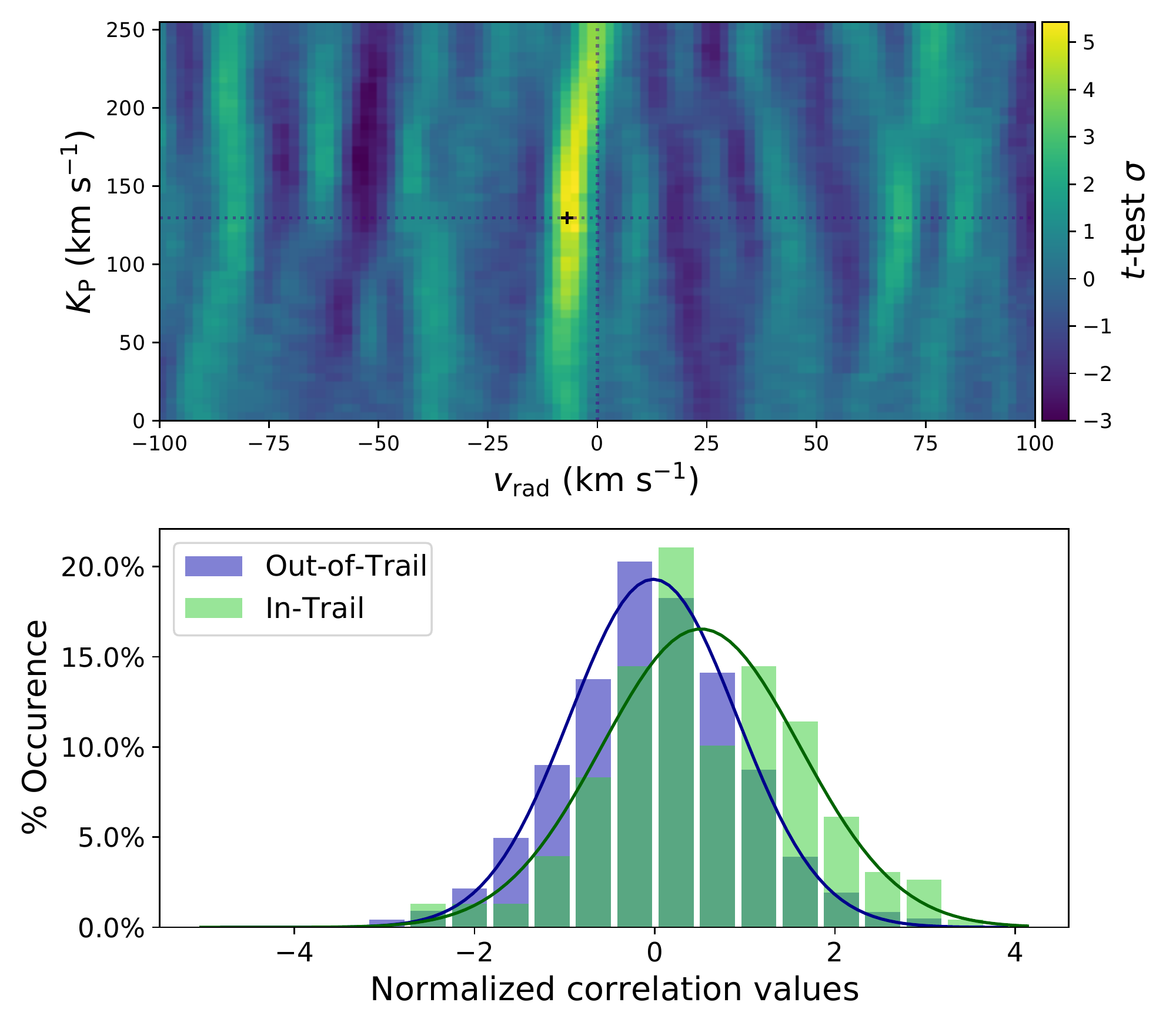}
\vspace{-0.5cm}
\caption{Welch \textit{t}-test results: (Top panel)  Scaled $t$-test detection significance as a function of \kp\ and \vvv{rad}. The \water\ signal is detected near the expected \kp\ value , but blue-shifted (the black cross indicates the peak position at $K_{\rm P, 0}$).
(Bottom panel) Distributions of cross-correlation values, normalized by the dispersion of the out-of-trail regions, away from (out-of-trail, blue), and near (in-trail, green) the planet RV (with $K_{\rm P, 0}$ and \vvv{rad}$=$\vpeakttestkpo\vpeakttestkpoerr\,\kms), with their associated best-fit Gaussian distributions, each with their corresponding mean and variance (blue and green lines, respectively). A detection of the transmission signal of the planet is expected to shift the in-trail distribution to higher correlation values, and this is what we observe, with the two distributions differing at the \sigttest$\,\sigma$ level. }
\label{fig:ttest_a2}
\end{figure}

The \kp\ versus \vvv{rad} $t$-test map is shown in Figure~\ref{fig:ttest_a2}. Note that to stay conservative, our $t$-test maps have been scaled such that there are no noise structures that are above the $\pm 3\sigma$ level. 
The $t$-test peak is found at \kp$ = $\kpttestmax\kpttestmaxerr\,\kms\ and \vvv{rad}$ = $\vradttestmax\vradttestmaxerr\,\kms, where the uncertainties correspond to a drop of 1$\sigma$ from the peak value. 
Large error bars on \kp\ are expected since the planet's acceleration during transit is low, especially as we only have partial transit sequences.
At $K_{\rm P,0}$, the signal is shifted at \vvv{rad}$ = $\vpeakttestkpo\vpeakttestkpoerr\,\kms, and the in-trail distribution is different from the out-of-trail one at the level of \sigttest$\sigma$, which highly supports our water detection. The CCF and \logl\ maps (\kp\ versus \vvv{rad} maps) are very similar, but are not shown to limit redundancy. The respective best-fit models for the CCF and \logl\ all lead to convincing detections even though their parameters are different, as also seen in \cite{Boucher2021}. Following the same arguments, we will put more credence on the model selection of the \logl, which will be used in the next section (Section~\ref{subsec:MCMC_a2}) for the full retrieval analysis.

In principle, the RME should be accounted for when molecular or atomic species are predicted to be present in both the stellar and planetary atmospheres. 
However, this effect is negligible since WASP-127 is a slow rotator ($v\sin (i_*) = 0.53^{+0.07}_{-0.05}$\kms; \citealt{Allart2020}). The resulting contamination signal would appear at $K_{\rm P} \simeq 4\,$km\,s$^{-1}$, which is far away from $K_{\rm P,0}$. Also, \water\ is not expected to be present in the atmosphere of a G5V star like WASP-127.

\begin{figure}
\centering
\includegraphics[scale=0.45]{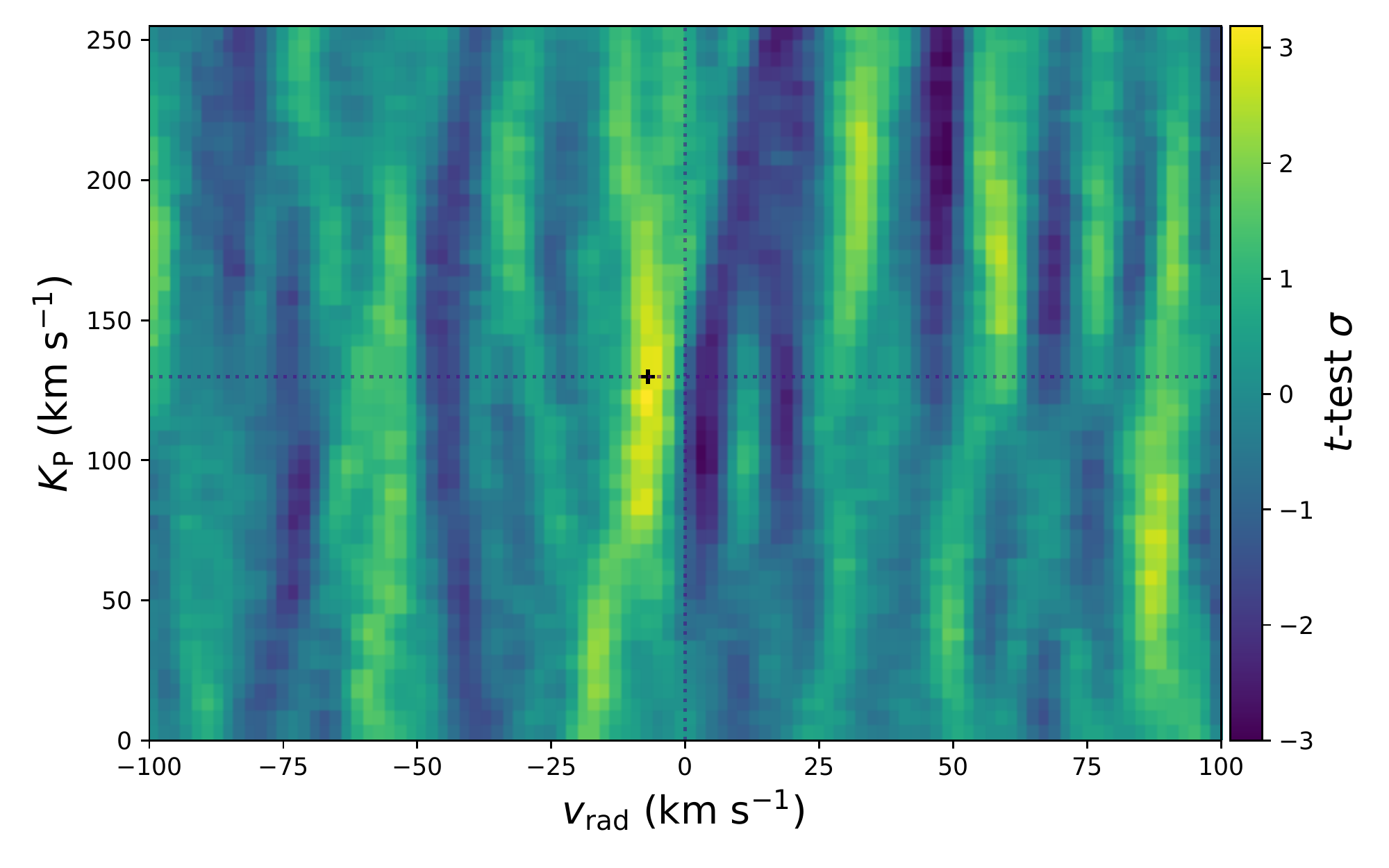}
\vspace{-0.5cm}
\caption{Same as Figure~\ref{fig:ttest_a2}, but for an OH-only model with a VMR $ = 10^{-5}$, $T_{\rm eq} = 800$\,K and \cloud$ = 0.1$\,bar. The \kp$_{,0}$ peak of \OHttest$\sigma$ is found at \OHv\,\kms, similar to the H$_2$O signal. }
\label{fig:ttest_OH}
\end{figure}

We also observe a tentative OH signal, at a scaled level of \OHttest$\sigma$, at a similar \vvv{rad}\ as for water (\vvv{rad} $ = $\OHv\,\kms, at \kp$_{,0}$), for a model only containing OH with a VMR $ = 10^{-5}$, $T_{\rm eq} = 800$\,K and \cloud$ = 0.1$\,bar. The other model parameters were set to be the same as for the best \water\ model. The resulting scaled $t$-test map is shown in Figure~\ref{fig:ttest_OH}. This finding is discussed further in Section~\ref{subsubsec:OH}.

\subsection{Full Retrievals}
\label{subsec:MCMC_a2}

We next computed full, free-parameter atmospheric retrievals.
We ran three types of retrievals: one using only the high-resolution SPIRou data (hereafter HRR, for High Resolution Retrieval), one using only the low-resolution HST WFC3, STIS and \emph{Spitzer} data from \citetalias{Spake2021} (hereafter LRR, for Low Resolution Retrieval), and finally, a joint retrieval by combining the low- and high-resolution data (hereafter JR, for Joint Retrieval). 
High-resolution models were generated from 0.9 to 2.55\,\um\ and low-resolution ($R=1000$) models from 0.3 to 5\,\um. Both model sets use the same values for all parameters. We down-sampled and binned the pre-existing high-resolution models (for JR) or the low-resolution models (for LRR) to $R = 75$ to match the HST WFC3 data, and $R = 50$ for STIS. Finally, we binned the low-resolution model over the \emph{Spitzer} IRAC1 (at 3.6\,\um) and IRAC2 (at 4.5\,\um) bandpass transmission functions (JR and LRR). 

Following \cite{Brogi2019}, we add the contribution of the low-dispersion spectroscopy (LDS) data to the log-likelihood using:
\begin{equation}
    \ln \mathcal{L}_{\rm tot} = \ln \mathcal{L}_{\rm HDS} + \ln \mathcal{L}_{\rm LDS},
\end{equation}
where $\ln \mathcal{L}_{\rm HDS}$ is given by equation~\ref{eq:loglikelihood_a2}, while $\ln \mathcal{L}_{\rm LDS}$ is given by:
\begin{equation}
    \ln \mathcal{L}_{\rm LDS} = -\frac{1}{2}\chi^2; 
\end{equation}
and the $\chi^2$ summation is done on the low-resolution data points.

For the HRR, we included the white-light transit depth in the \logl\ computation. Namely, we made use of the down-sampled model, compared its mean transit depth to the mean of the WFC3 data, and added that to the \logl. This was done to anchor the transit depth to its observed value and to limit the exploration in the parameter space that would otherwise lead to completely different transit depths\footnote{The exclusion of the white-light transit depth in the \logl\ calculation led to much higher radius and much lower temperature, both of which affect the line contrast in a correlated manner. }, since all continuum information is lost in the analysis of the high-resolution data.

We included the opacity contribution from \water, CO, \diox, FeH, \methane, HCN, NH$_3$, C$_2$H$_2$, TiO, OH\footnote{We ran retrievals that did include OH, but also retrievals that did \emph{not} include OH. Their differences are discussed in \ref{subsubsec:OH}.}, as well as Na and K for completeness and to achieve a better fit to the STIS data. 
While most included species are not necessarily expected to be detected (based on the sensitivity of the SPIRou data, see Appendix~\ref{subsec:injec-recov}), useful constraints on their abundances may still be obtained and rule out certain chemical scenarios.
This is also crucial to properly estimate elemental abundance ratios (C/O, C/H, O/H), and to limit the biases that would come from not including all the relevant molecules. 
We let $T_{\rm eq}$ and $R_{\rm P}$ vary, while the reference pressure is still fixed to $P_0 = 10$\,mbar. With the inclusion of the STIS data covering shorter wavelengths, we also include the two scattering parameters, i.e. the scattering index $\gamma$ and the common logarithm of the enhancement factor $S$ from Equation~\ref{eq:scat}, as free parameters. 
The sixteen (seventeen, when including OH) model parameters are thus: the VMRs of \water, CO, \diox, FeH, \methane, HCN, NH$_3$, C$_2$H$_2$, TiO, , Na and K, \tp\ (K), $\log_{10}$ \cloud\ (bar), $R_{\rm P}$ ($R_{\rm J}$), $\gamma$, and $\log_{10} S$. 
To these, we added the following orbital and dynamical parameters: \kp, \vvv{rad}, $\omega$ (d\,$^{-1}$), and $f$, the last two being the solid rotation angular frequency and the intensity fraction of the right hemisphere (morning side) of the kernel. 
Several of these parameters will be sensitive only when either the low-resolution or the high-resolution data are included, but should not alter the results or lead to scenarios that have no physical interpretations when they are included even without sensitivity.  
Our atmospheric retrievals therefore include a total of 20 (21, with OH) parameters.


We used a Markov Chain Monte Carlo (MCMC) to explore the parameter space and compute the posterior distribution of each parameter and estimate their uncertainties. The MCMC sampling was done using the \texttt{python} library \texttt{emcee} \citep{emcee}, which implements the affine-invariant ensemble sampler by \cite{Goodman2010}.
For each of our runs, we combined the three transit sequences, included all orders, and followed 64 walkers until convergence.

Once a steady state was reached, we ran 8900, 31500, and 8900 steps for the HRR, LRR, and JR, respectively, i.e. where we had at least 10 times the auto-correlation length in our amount of steps. The priors that were used are listed in Table~\ref{tab:param_MCMC_a2}. The priors for all parameters, except \kp, are uniform. The \kp\ prior was chosen to be a Gaussian centered on the previously measured value of $K_{\rm P} = 129.95$\,\kms with $\sigma = 1.2\,$\kms. For completeness, we tested free \kp\ JR after the fact, and found that it does not significantly affect the results. 

\begin{figure*}
\vspace{3.0cm}
\begin{picture}(400,400)
\put(-105,-50){\includegraphics[scale=0.53]{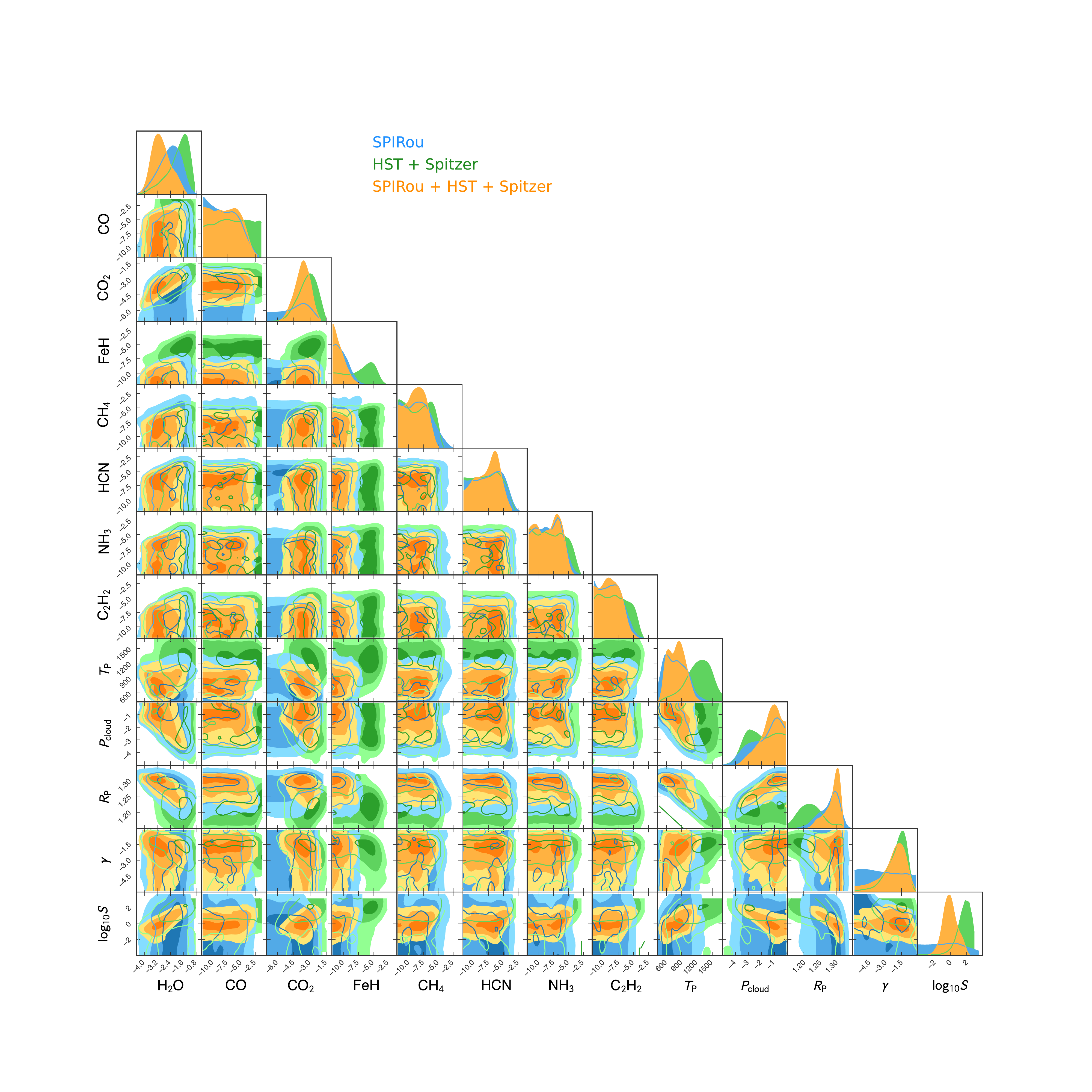}}
\put(260,350){\includegraphics[scale=0.43]{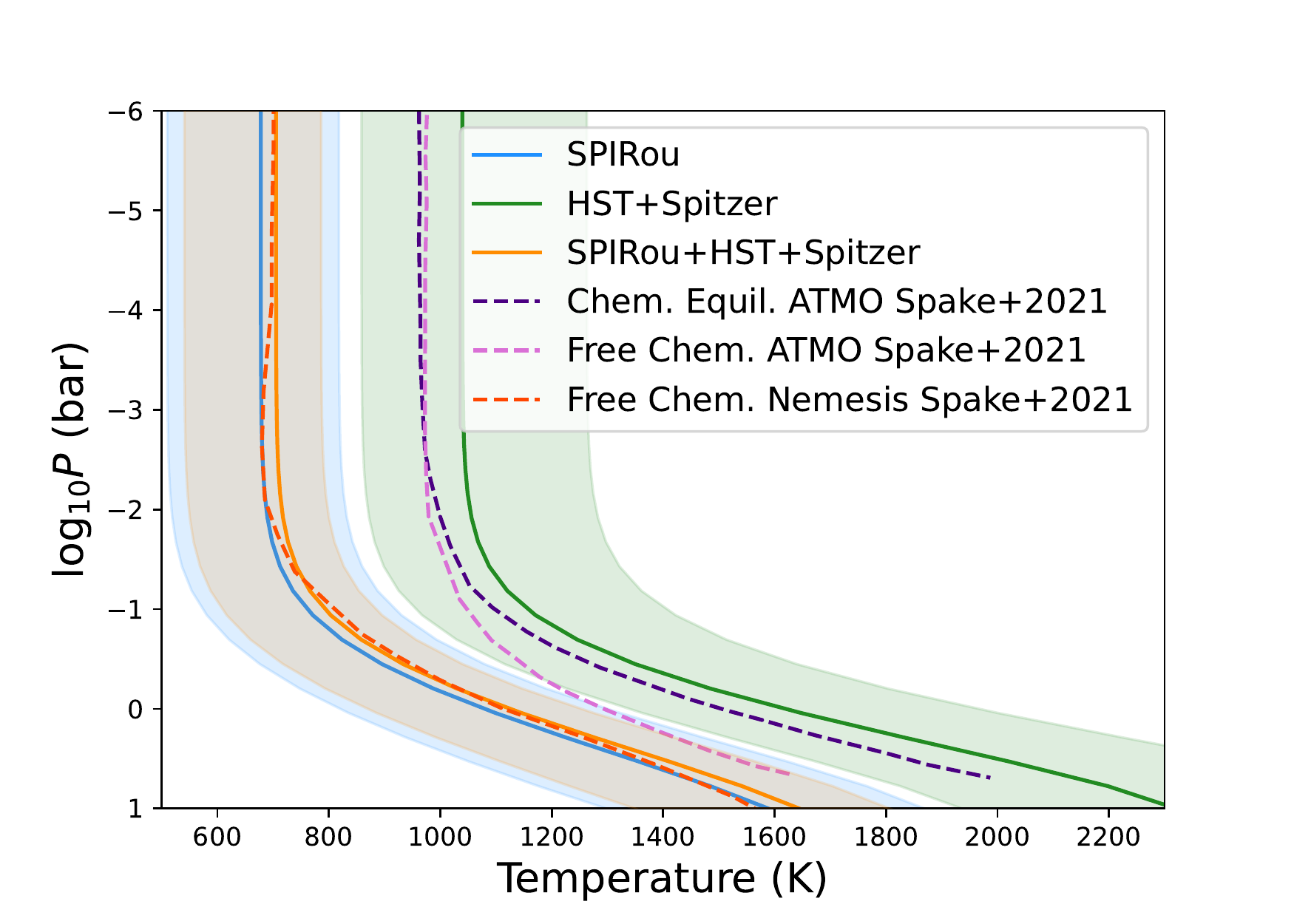}}
\put(258,210){\includegraphics[scale=0.40]{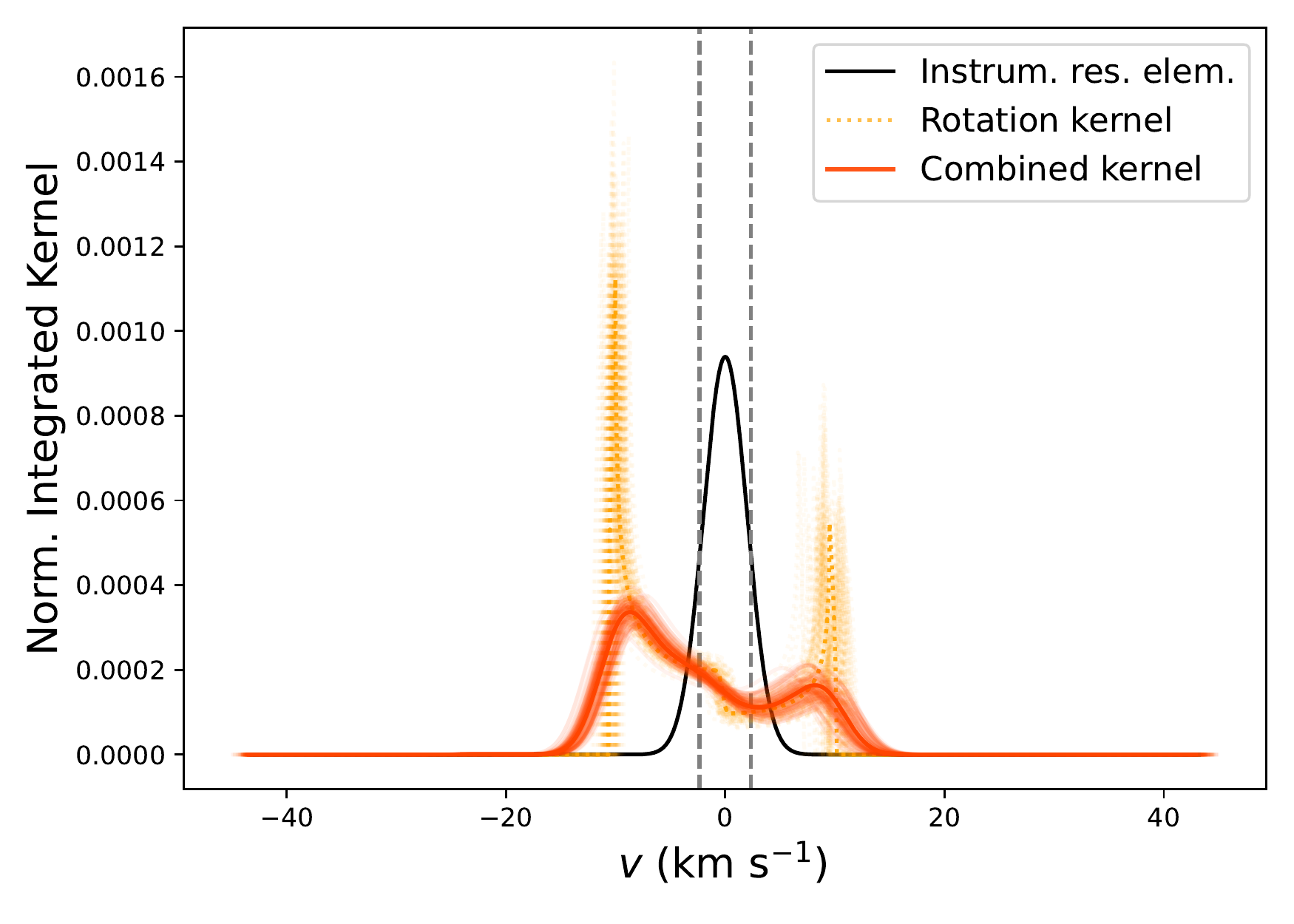}}
\end{picture}
\caption{ \small
Constraints on the atmospheric properties retrieved with the three SPIRou transits and/or the \citetalias{Spake2021} HST and \emph{Spitzer} data of \waspb. (Bottom left) Posterior probability distributions for the parameters of the MCMC fits using a parameterized T-P atmosphere with free chemical abundances, a gray cloud deck and hazes, and Gaussian rotational broadening. The results for the high-resolution only (HRR) retrieval are shown in blue, the low-resolution only (LRR) is in green, and the joint retrieval (JR) is shown in orange. The posterior distributions of TiO, Na, K, \kp, \vvv{rad}, $\omega$ and $f$ are not shown to improve visibility, and because they do not affect our conclusions. 
(Top right) Median retrieved Guillot T-P profiles where the shaded region represents the $1\,\sigma$ uncertainties, compared to the mean profiles retrieved in \citetalias{Spake2021} (dashed lines). 
The median retrieved profiles have a temperature of 
$1079^{+220}_{-260}$\,K, $1618^{+347}_{-281}$\,K, and $1125^{+128}_{-256}$\,K, 
for HRR, LRR and JR respectively, at 1\,bar. The HRR and JR are consistent with the NEMESIS profile from \citetalias{Spake2021}, and the LRR to the ATMO ones, but they are all consistent with one another within $\sim$2\,$\sigma$.
(Middle right) Retrieved rotation kernel for the JR. The instrument profile is shown in black and 100 random transit solid rotation kernels are shown in orange (dotted curves), with their respective combined profile in red (plain curve). The resulting kernel parameters converge to \vvv{rad} \vpeakJR\vpeakJRerr\,\kms, $\omega = $\fwhmJR\,d$^{-1}$ ($\approx 9.8^{+0.5}_{-0.5}$\,\kms at the equator), and $f= 0.49^{+0.09}_{-0.10}$.
}
\label{fig:mcmc_a2}
\end{figure*}

\begin{figure*}
\includegraphics[width=\textwidth]{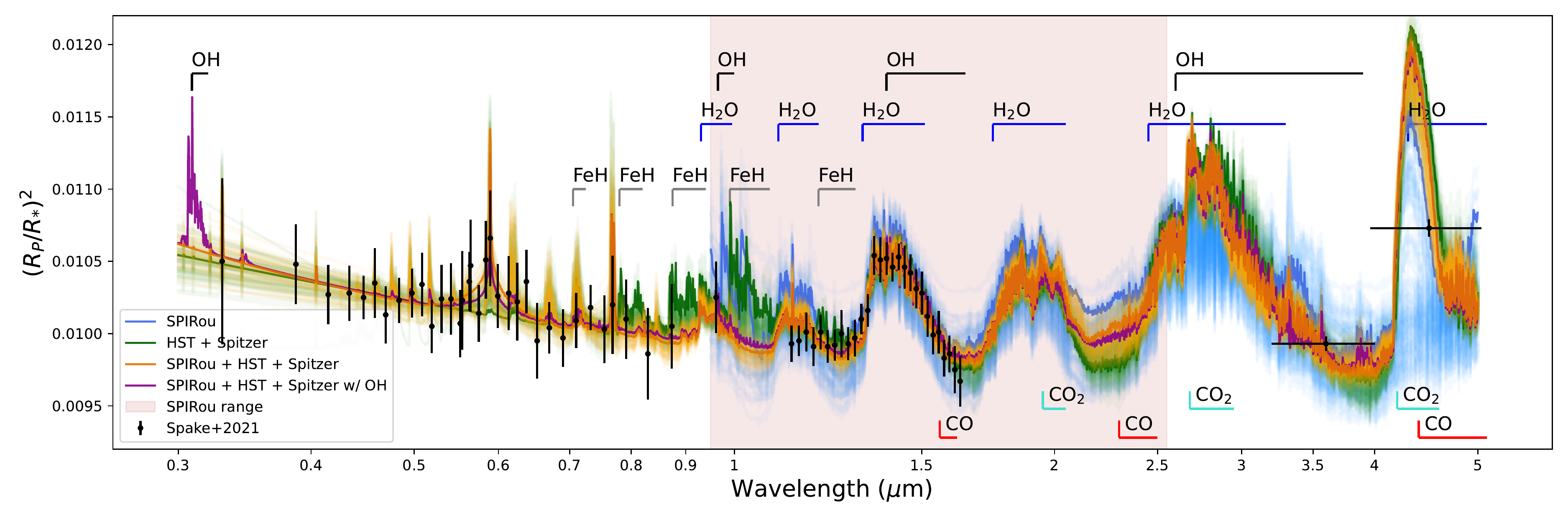}
\vspace{-0.5cm}
\caption{  
Best \logl\ (dark) and 100 randomly generated from MCMC sample models (light) of the transmission spectrum of WASP-127\,b (with $R=1000$) using the HRR (blue), the LRR (green) and the JR parameters (orange), compared to the HST and \emph{Spitzer} data from \citetalias{Spake2021} (black). 
The best-fit model with OH is shown in magenta for comparison. 
The red shaded region shows the wavelength range of SPIRou. 
The position of the absorption bands of the major molecules considered are shown. 
}
\label{fig:best-fit-model}
\vspace*{0.1cm}
\end{figure*}

The resulting posterior probability distributions for our three types of retrieval (HRR, LRR and JR), their associated median T-P profiles, and the retrieved shape of the rotation kernel for the JR are shown in Figure~\ref{fig:mcmc_a2}. Their respective median parameters are also tabulated in Table~\ref{tab:param_MCMC_a2} along with their 1\,$\sigma$ uncertainties, corresponding to the range containing 68\% of the MCMC samples, or their 2\,$\sigma$ upper or lower limit (for non-detections, at 95.4\%). Though all 20 (21 with OH) parameters were included in all of the fits, we chose to remove some from the corner plots to improve visualization --- namely those which showed less ``relevant'' non-detections and/or no correlation with other parameters. These parameters are TiO, Na, K, \kp, \vvv{rad}, $\omega$, and $f$. 
The best-fit models (using the set of parameters yielding the highest \logl\ for each type) are shown in Figure~\ref{fig:best-fit-model}, and all provide a generally good fit to the \citetalias{Spake2021} data.

\subsubsection{Joint Retrieval}
\label{subsubsec:JR}

The differences between our three types of retrievals (HRR, LRR and JR) are discussed in Section~\ref{subsec:diff_retrievals}, but from here onward, we will focus on the JR as it provides the best constraints on all parameters. Additionally, we present the results from the retrievals without OH, as its presence remains uncertain, but it is further discussed in Section~\ref{subsubsec:OH}.

We measure solar\footnote{The solar abundances for a chemical equilibrium \texttt{FastChem} model of \waspb, with the retrieved JR $R_{\rm P}$ and T-P profile, are $\log_{10}$\water\ $ \simeq -3.4$, $\log_{10}$CO $ \simeq -3.3$, and $\log_{10}$\diox\ $ \simeq -6.0$.} \water, but super-solar \diox\ abundances of $\log_{10}$\water\ $= $ \waterJR\waterJRerr\ and $\log_{10}$\diox\ $= $ \dioxJR\dioxJRerr. This represents values that are 0.6 to $7\times$ solar for \water, and roughly 50 to up to $1200\times$ solar for \diox\footnote{The results are similar if we include the contribution of OH, but with a decrease of the median abundance to $-3.3$ for \water\ ($\sim 1.5\times$ solar) and $-4$ for \diox ($\sim 100\times$ solar).}. 
We note that our constraint on the \diox\ abundance is mainly driven by the \emph{Spitzer} IRAC2 data point, but that its upper limit from the HRR is comparable to the volume mixing ratio from the JR. 
We retrieve upper limits of $\log_{10}$CO $ \lesssim $ \COJR\ (2$\sigma$ limit, which is sub-solar and equivalent to $< 0.2\times$ solar; discussed further in Section~\ref{subsubsec:CO2}), and $\log_{10}$FeH $ \lesssim \FeHJR$\ ($2\sigma$, discussed further in Section~\ref{subsubsec:FeH}). 
The \diox\ detection combined with the upper limit found for CO excludes the \citetalias{Spake2021} chemical equilibrium scenario. Since a CO abundance greater than $\log_{10}$CO$ \simeq -4$ to $-3$ (depending on the cloud/hazes conditions) would have been detected with SPIRou only, not seeing it in both retrievals which used SPIRou (HRR and JR) means that the signal at 4.5\,\um\ comes (in large part) from \diox. 

\begin{table*}
\caption{MCMC Retrieval Parameter Priors and Results \label{tab:param_MCMC_a2}}
\begin{threeparttable}
\begin{tabular}{lcrrrc}
\hline
Parameter  & Priors & SPIRou only & HST+Spitzer only  & SPIRou+HST+Spitzer  & Unit \\
\hline
$\log_{10}$ \water & $\mathcal{U}(-12,-0.1)$              & $-2.39_{-0.57}^{+0.73}$  &  $-1.80_{-0.42}^{+0.68}$  &  $-3.02_{-0.63}^{+0.47}$ &  \\ 
$\log_{10}$ CO & $\mathcal{U}(-12,-0.1)$                  & $< -3.85$  &  $< -1.25$  &  $< -4.04$ &  \\ 
$\log_{10}$ \diox & $\mathcal{U}(-12,-0.1)$               & $-6.42_{-2.42}^{+3.70}$  &  $-3.15_{-0.70}^{+0.94}$  &  $-3.66_{-0.61}^{+0.76}$ &  \\ 
$\log_{10}$ FeH & $\mathcal{U}(-12,-0.1)$                 & $< -8.19$  &  $< -4.14$  &  $< -8.73$ &  \\ 
$\log_{10}$ \methane & $\mathcal{U}(-12,-0.1)$            & $< -4.78$  &  $< -5.19$  &  $< -6.23$ &  \\ 
$\log_{10}$ HCN & $\mathcal{U}(-12,-0.1)$                 & $< -4.09$  &  $< -4.17$  &  $< -4.82$ &  \\ 
$\log_{10}$ NH$_3$ & $\mathcal{U}(-12,-0.1)$              & $< -5.20$  &  $< -4.24$  &  $< -5.57$ &  \\ 
$\log_{10}$ C$_2$H$_2$ & $\mathcal{U}(-12,-0.1)$          & $< -5.79$  &  $< -4.63$  &  $< -6.27$ &  \\ 
$\log_{10}$ TiO & $\mathcal{U}(-12,-0.1)$                 & $< -3.11$  &  $< -7.15$  &  $< -8.85$ &  \\ 
$\log_{10}$ Na & $\mathcal{U}(-12,-0.1)$                  & $\cdots$  &  $-4.86_{-2.04}^{+3.37}$  &  $-6.59_{-1.85}^{+2.02}$ &  \\ 
$\log_{10}$ K & $\mathcal{U}(-12,-0.1)$                   & $\cdots$  &  $< -2.64$  &  $< -5.26$ &  \\ 
$T_{\rm P}$ & $\mathcal{U}(400, 2000)$                    & $769_{-253}^{+180}$  &  $1257_{-215}^{+266}$  &  $800_{-229}^{+100}$ & K \\ 
$\log_{10} P_{\rm cloud}$ & $\mathcal{U}(-6.0, 0.0)$      & $> -3.69$  &  $> -3.67$  &  $> -2.88$ & bar \\ 
$R_{\rm P}$ & $\mathcal{U}(1.0, 1.6)$                     & $1.29_{-0.03}^{+0.03}$  &  $1.22_{-0.04}^{+0.03}$  &  $1.30_{-0.01}^{+0.02}$ & $R_J$ \\ 
$\gamma$ & $\mathcal{U}(-6.0, 0.0)$                       & $\cdots$  &  $-1.75_{-0.65}^{+0.89}$  &  $-2.04_{-0.89}^{+1.11}$ &  \\ 
$\log_{10} S$ & $\mathcal{U}(-10.0, 4.0)$                 & $\cdots$  &  $1.94_{-0.54}^{+0.72}$  &  $-0.07_{-0.83}^{+0.87}$ &  \\ 
$K_{\rm P}$ & $\mathcal{G}(129.95, 1.2)$                  & $130.21_{-1.15}^{+1.16}$  &  $\cdots$  &  $130.44_{-1.19}^{+1.13}$ & km s$^{-1}$ \\ 
$v_{\rm rad}$ & $\mathcal{U}(-10.0, 10.0)$                & $-0.28_{-0.67}^{+0.57}$  &  $\cdots$  &  $-0.26_{-0.65}^{+0.67}$ & km s$^{-1}$ \\ 
$\omega$ & $\mathcal{U}(0, 15)$                           & $9.17_{-0.53}^{+0.45}$  &  $\cdots$  &  $9.10_{-0.51}^{+0.49}$ & d$^{-1}$ \\ 
$f$ & $\mathcal{U}(0, 1.0)$                               & $0.52_{-0.12}^{+0.10}$  &  $\cdots$  &  $0.49_{-0.10}^{+0.09}$ &  \\ 
\hline
 C/O  & $\cdots$                                          & $0.02_{-0.02}^{+0.01}$  &  $0.04_{-0.03}^{+0.08}$  &  $0.10_{-0.06}^{+0.10}$ &  \\ 
$[$C/H$]^a$ & $\cdots$                                    & $-0.55_{-1.87}^{+0.94}$  &  $0.57_{-0.86}^{+0.83}$  &  $-0.26_{-0.67}^{+0.69}$ &  \\ 
$[$O/H$]^a$ & $\cdots$                                    & $0.84_{-0.64}^{+0.61}$  &  $1.46_{-0.45}^{+0.54}$  &  $0.21_{-0.62}^{+0.56}$ &  \\
\hline
\end{tabular}

\begin{tablenotes}
\item \textsc{\textbf{Notes}} --- The marginalized parameters from the likelihood analysis with their $\pm 1\,\sigma$ errors, or their 2$\sigma$ upper or lower limits. The log of the VMR are unitless. $^a$Abundances ratios compared to the solar value in $\log_{10}$.
\end{tablenotes}
\end{threeparttable}
\end{table*}

\begin{figure}
\includegraphics[width=\linewidth]{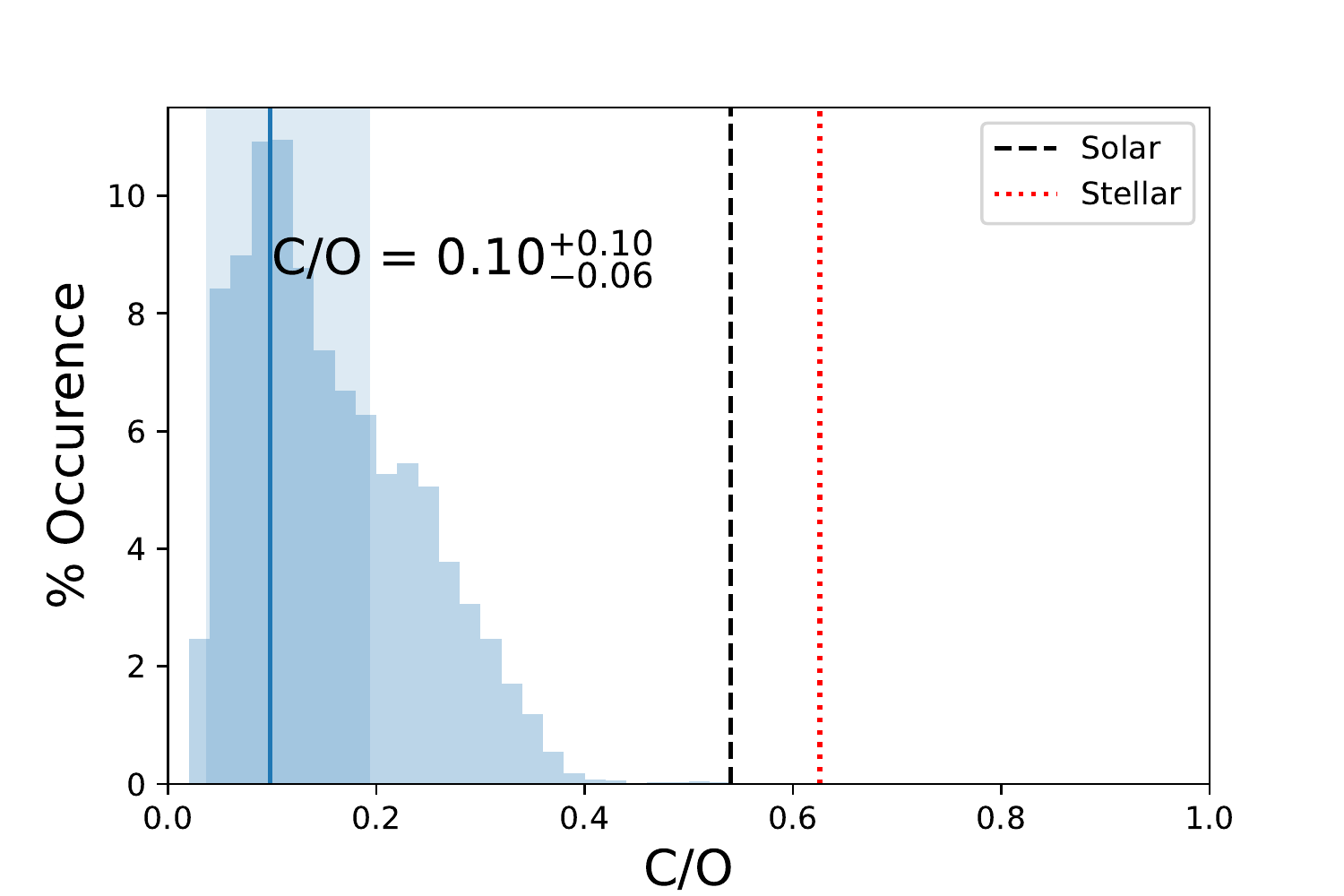}
\caption{  
Posterior distribution of the C/O ratio for the JR retrievals with and without OH. The distribution is shown for the retrieval without OH. 
We find that it points toward sub-solar/stellar, with C/O $\lesssim$ \CsurOupperlimtwosig\ ($2\sigma$), as seen from comparing to the computed values of solar (black dashed line) and the stellar-composition (red dotted line).
}
\label{fig:best-fit-c_sur_o}
\vspace*{0.1cm}
\end{figure}

\begin{figure}
\includegraphics[width=\linewidth]{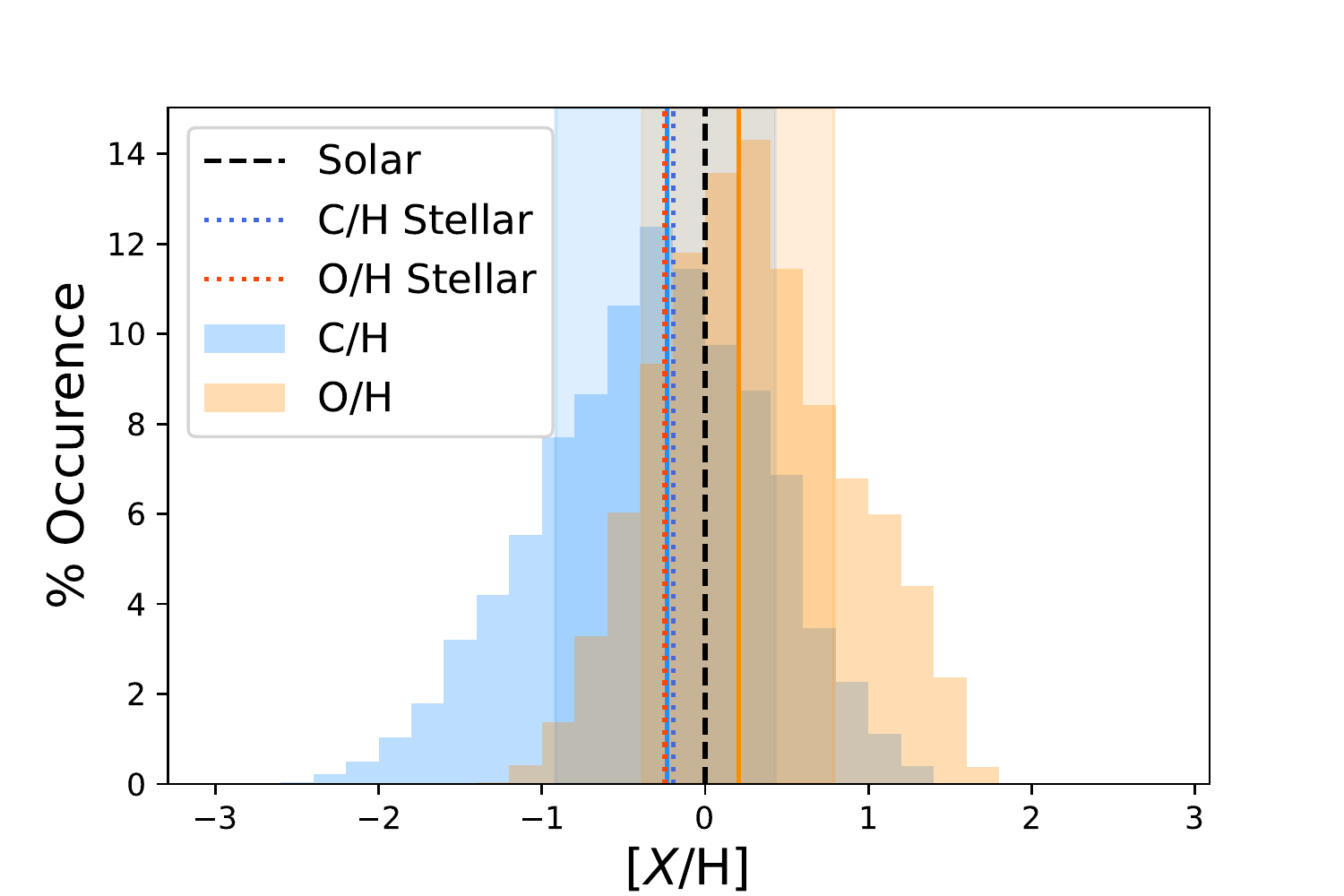}
\caption{  
Distributions of the [C/H] (blue), and [O/H] (orange) ratios of the JR, i.e. the $\log_{10}$ of abundance ratios compared to the solar values. The color-matched lines show the maximum of the distribution and the shaded region shows the 1$\sigma$ uncertainties. The black dashed and colored dotted (blue and red) lines shows the solar and stellar ratios, respectively (assuming that the stellar atomic C and O abundances follow [M/H]$ = -0.193$).  
}
\label{fig:best-fit-abund}
\vspace*{0.1cm}
\end{figure}


We also obtain non-detections for \methane, C$_2$H$_2$, TiO and K (consistent with  \citealt{palle_feature-rich_2017}, \citealt{Allart2020}, and \citetalias{Spake2021}). HCN and NH$_3$ are also not detected (their upper limits are listed in Table~\ref{tab:param_MCMC_a2}), but the retrieval slightly favors models including some, which we discuss briefly in Section~\ref{subsubsec:HCN}. We recover the Na detection, $\log_{10}$Na $ =$\NaJR\NaJRerr, which is roughly consistent with the retrieved value in \citetalias{Spake2021} ($\simeq -7.0 \pm 0.4$), but with much larger error bars, probably due to our inclusion of a cloud deck.

The T-P profile (Figure~\ref{fig:mcmc_a2}) is consistent with the NEMESIS retrieval from \citetalias{Spake2021}, but at lower temperature than their ATMO retrieval. The cloud top pressure is located at $\log_{10}$\cloud$= -1.0^{+0.9}_{-0.6}$\,bar, which is lower in altitude than, but still consistent (within $2\sigma$) with what was found by \cite{Allart2020} and \cite{Skaf2020}; both roughly around $10^{-3}$\,bar. For \cite{Skaf2020}, the low-resolution of their observations could have hampered a precise determination of \cloud, and for \cite{Allart2020}, the slight difference could come from the larger value of $R_{\rm P}$ that they used: a larger planet radius would need higher clouds to dampen a given signal amplitude. Still, the LRR and HRR \cloud\ are in line with the literature values, which seems to indicate that it is driven by the low-resolution data or the lack of continuum at high-resolution. Comparatively, in \citetalias{Spake2021}, they did not include a grey cloud deck contribution as they did not find evidence in the data to support its presence, but rather a stronger haze signal.

On that matter, we observe that our retrieved radius of $R_{\rm P} \simeq 1.30\,R_{\rm J}$ (for $P_0 = 10$\,mbar) is much smaller than the ATMO values in \citetalias{Spake2021} ($R_{\rm P}$ between 1.38 and 1.45\,$R_{\rm J}$ at 1\,bar, roughly equivalent to $\sim 1.53$--$1.61\,R_{\rm J}$ at $P_0$), but is more consistent with their NEMESIS radius ($1.30 \pm 0.2 \,R_{\rm J}$), and also with that of \cite{Skaf2020} ($R_{\rm P} \simeq 1.16\,R_{\rm J}$ at 10\,bar, equivalent to $1.32\,R_{\rm J}$ at $P_0$). It is also in line with the value from \cite{Seidel2020_wasp127}, at $1.30 \pm 0.2\,R_{\rm J}$, from which we took most of our fixed parameters values. 
Even though $R_{\rm P}$ is degenerate with the chemical abundance, \cloud, and temperature, our constraint on the radius is extremely tight due to the combination of the low- and high-resolution data, which will be discussed further in Section~\ref{subsec:diff_retrievals}.  

Uncorrelated/non-degenerate posterior distributions are found (not shown) for the shift of the planet signal\footnote{We observe a very slight double-peaked correlation between \vvv{rad} and $\omega$, but their marginalized posterior distributions remains single-peaked and well constrained.}, \vvv{rad} $ =$ \vpeakJR\vpeakJRerr\,\kms, the solid rotation frequency, $\omega = $ \fwhmJR\,d$^{-1}$ ($\approx 9.8^{+0.5}_{-0.5}$\,\kms at the equator), and the morning terminator kernel intensity fraction, with $f = $ \cloudfracJR. 
The shape of the kernel is well constrained (as seen on Figure~\ref{fig:mcmc_a2}, middle right panel) and suggests a large broadening, with a dampened morning terminator (right side). This large broadening behavior was observed in all retrieval runs attempted here, as well as in prior runs where a Gaussian kernel was used and which recovered the same large blue-shift as seen in the $t$-test maps (Figures~\ref{fig:CCF_2d_a2}--\ref{fig:ttest_OH}). 
This blue shift can thus be reproduced by having a rapidly rotating planet or atmosphere and no additional radial velocity shift. This is much larger than the expected synchronous rotation rate of \waspb\ that is only $\sim 1.6$\,\kms, but super-rotation could amplify the broadening. A puffy atmosphere like \waspb's could have interesting dynamics and 3D models would certainly help shed light on these results.
The rotation kernel and radial velocity shift are discussed in more details in Section~\ref{subsec:wind_a2}. 

Looking back at Figure~\ref{fig:mcmc_a2}, we see that the abundances of some molecules are correlated with \cloud, which results in broader constraints on absolute abundances. 
However, the abundance posteriors of some molecules are correlated (especially \water\ and \diox) due to their similar dependencies on the continuum level, which means that their relative abundances can be accurately constrained \citep{Gibson2022}. 
The \water/\diox\ ratio is constrained to $\log_{10}$\water/\diox\ $ = $ \waterdiox\waterdioxerr, i.e., \water\ is between $\sim$ 2 and 12 times more abundant than \diox\ in \waspb's atmosphere.

We can also combine the retrieved constraints of all carbon- and oxygen-bearing molecules to compute the atmospheric C/O ratio as:

\begin{equation}
    \mathrm{C/O}_{\mathrm{gas}} = \frac{ \mathrm{CO} + \mathrm{CO}_2 + \mathrm{CH}_4 + \mathrm{HCN} + 2 \mathrm{C}_2\mathrm{H}_2}{ \mathrm{H}_2\mathrm{O} + \mathrm{CO} + 2 \mathrm{CO}_2 + \mathrm{TiO}},
    \label{eq:csuro}
\end{equation}
assuming that no other major C- or O- bearing molecules are present. We chose to compute the elemental ratios excluding OH as its abundance may be overestimated (see Section~\ref{subsubsec:OH}), which in turn could bias \waspb's inferred atmospheric O/H and C/O ratios. 
We computed C/H and O/H in a similar fashion, but using H = $2 \mathrm{H}_2\mathrm{O} + 4 \mathrm{CH}_4 + \mathrm{HCN} + 3 \mathrm{NH}_3 + 2 \mathrm{C}_2\mathrm{H}_2 + 2 (1-\sum_i X_i) \mathrm{H}_2$, where the last term represents twice the total abundance of H$_2$, i.e., 0.85 scaled by the total minus the sum of the other molecules. For comparison, we generated \texttt{FastChem} chemical equilibrium models of planetary atmospheres with solar\footnote{The solar abundance from \texttt{FastChem} are based on \cite{Lodders2009_solar_abund}} and stellar compositions (where we assumed that the stellar C, N, and O abundances followed the estimated metallicity trend of [M/H]$=-0.193$; \citealt{Stassun2019}), and using the retrieved JR T-P profile and $R_{\rm P}$. We took the median abundance over the pressure region between $10^{2}$ and $10^{-6}$\,bar, and then computed the atomic ratios with the same equation \ref{eq:csuro}. From those abundances, we obtain solar and stellar composition C/O equal to 0.55 and 0.63, respectively, while the $\log_{10}$ C/H and O/H are respectively $-3.56$, and $-3.31$ for the solar composition mix, and $-3.76$, and $-3.55$ for stellar composition. The resulting posterior of the C/O for WASP-127\,b is shown in Figure~\ref{fig:best-fit-c_sur_o}, the C/H and O/H distributions are shown in Figure~\ref{fig:best-fit-abund}, and their best-fit values are listed in Table~\ref{tab:param_MCMC_a2}. 

We obtain a highly sub-solar C/O of \CsurOdistnoOH\CsurOdisterrnoOH, compared to the computed C/O$_{\odot} = 0.55$ (for the solar composition mix, close to the common literature value of $0.54$, as expected), and the computed stellar composition\footnote{To be more specific, these solar and stellar composition abundance ratios are computed using the same sub-sample of molecules that are used to compute our retrieved \waspb\ values. } one of C/O$_{*} = 0.63$. This indicates that the atmosphere of \waspb\ is either oxygen-rich or carbon-depleted, or both, compared to what it should be with solar or stellar composition. A solar value is rejected at more than $4\sigma$. 
On the contrary, within their uncertainties, both the C/H and O/H ratios are consistent with being solar/stellar. 
The choice of which molecules are included in these ratio calculations can significantly change the inferred results. Nonetheless, since we computed the expected planetary atmosphere abundance ratios for solar and stellar compositions in the same way as for our actual results, it should be a good means of comparison. 
The formation and evolution mechanisms that could reproduce such elemental abundance ratios in \waspb's atmosphere are explored in Section~\ref{subsec:formation}.

\section{Discussion}
\label{sec:Discuss_a2}

The primary goal of this study was to differentiate between the two \citetalias{Spake2021} scenarios --- a goal we accomplished by showing that 
\waspb\ does not have a CO abundance high enough to explain the \citetalias{Spake2021} \emph{Spitzer} results (expected to be around $\log_{10}$CO $\simeq -1.9$ for their chemical equilibrium scenario), but rather has $\log_{10}$CO $< $ \COJR. We were able to set a constraint on the C/O ratio at C/O $ = $ \CsurOdistnoOH\CsurOdisterrnoOH. 
Finally, we detect water at a very high significance, and potentially detect a signal consistent with OH. These partially unexpected results should be put in context to be better understood. 
In this section, we will first discuss some of the molecular detections and non-detections that were obtained. A comparison between the different types of retrievals will follow. Then, we will briefly review the observed shift and broadening. Finally, we will explore the potential formation mechanisms of \waspb.

\subsection{Molecular (non-)detections}
\label{subsec:detections}

\subsubsection{CO and CO$_2$}
\label{subsubsec:CO2}
The non-detection of CO is surprising because, as stated in \cite{Spake2021}, a scenario where all of the carbon is found in \diox\ and little to none in CO would be thermochemically implausible as no obvious non-equilibrium mechanism would deplete CO by many orders of magnitude while enhancing \diox. 
The relative abundance between \diox\ and CO has been explored in \cite{Heng2016_co2}. They conclude that \diox\ should rarely be dominant compared to CO (\diox/CO $\simeq 1$) and \water\ in hot and H$_2$ dominated atmospheres, and that it can be dominant if the C abundance is enhanced by a few orders of magnitude compared to solar, which is not our case \citep[see also][]{moses_compositional_2013, line_systematic_2014}. 
As mentioned above, we get a strong constraint on \diox, but this detection and its retrieved abundance ($\sim -3.7$; which could still be sub-dominant to CO given its upper limit at $\sim $ \COJR, within the uncertainties) are highly reliant on the IRAC2 4.5\,\um\ point. 
Still, the best-fit HRR model (SPIRou only) favors the inclusion of \diox\ (with a $\log_{10}$ VMR $\sim -3.8$). 
A strong \diox\ signal was recently detected in the JWST NIRSpec PRISM and G395H data of WASP-39\,b from the JWST Early Release Science program. The analyses agree on WASP-39\,b having a super-solar metallicity (around 3 to 10$\times$ solar), but some studies retrieve sub-solar to solar C/O \citep[roughly between 0.3 to 0.45;][]{JWST2022_CO2, Alderson2023}, while others get super-solar C/O \citep[between 0.6 to 0.7;][]{Rustamkulov2022}. Such differences highlight the difficulty that may arise when trying to uncover formation and evolution pathways. This is discussed further in Section~\ref{subsec:formation}.

\subsubsection{OH}
\label{subsubsec:OH}

Our results indicate a marginal OH signal, right at the limit of significance according to our scaled $t$-test (\OHttest$\sigma$). The OH signal is highly visible in Tr2, partially visible in Tr1, but not visible in Tr3 (which has relatively poorer data quality). 
Overall, we consider this detection as tentative and it will necessitate future confirmation. From a pure equilibrium chemistry standpoint, OH is not expected to be present on \waspb\ as it would normally imply \water\ dissociation, and \waspb's equilibrium temperature of 1400\,K is too low for such dissociation to occur (\water\ dissociation is expected to occur for $T_{\rm dayside} \gtrsim 2500\,$K; \citealt{parmentier_thermal_2018}). 
The uncertainty of our OH detection motivated our runs of retrievals that excluded its contribution, but we still ran some that included OH. 
From those, we obtain a surprisingly good constraint on the abundance, at $\log_{10}$OH $ =$\OHJR\OHJRerr\ in the JR, and goes up to $-2.48_{-0.54}^{+0.67}$ in the HRR, but remains unconstrained in the LRR.  
We find that this inclusion affects a little the retrieved abundance ratios, but not enough to change our conclusions.
The value increases to $\log_{10}$ O/H $= -2.9 \pm 0.6$ ($\simeq 2.6\times$ solar $\simeq 4.5\times$ stellar), compared to its median value of $-3.1 \pm 0.6$ ($\simeq 1.6\times$ solar $\simeq 2.9\times$ stellar), when excluding the OH.
The $\log_{10}$ C/H decreases to a value of $-4.2 \pm 0.7$ ($\simeq 0.2\times$ solar $\simeq 0.3\times$ stellar), compared to its median value of $-3.8 \pm 0.7$ ($\simeq 0.6\times$ solar $\simeq 0.9\times$ stellar), when excluding the OH, which puts it in a slightly more sub-stellar state. This also decreases the C/O to $\simeq 0.04\inc{0.04}{0.03}$, which is even more sub-solar.
With the highest \logl\ model parameters including OH (magenta curve on Figure~\ref{fig:best-fit-model}), 
we re-computed the scaled $t$-test and found a value of $6.5\sigma$; indicating that the inclusion of the other molecules and other fitted parameters is beneficial ($+1.2\sigma$ compared to just having water). Comparatively, the best-fit JR model optimized without OH (that has slightly higher abundances of \water\ and \diox, and slightly larger $R_{\rm P}$) yields smaller CCF S/N and $t$-test (decreased by $\sim 0.6$\,dex and $\sim 1.0\sigma$, respectively). However, it results in a insignificant drop of \logl, with a $\Delta$BIC $\simeq 0$. This means that our analysis is unable to unambiguously differentiate between the two scenarios. 
We note that the degeneracy between $R_{\rm P}$ and the abundance could have an impact here. If the OH is indeed present and comes from higher regions in the atmosphere, the equivalent radius should be greater. The fact that it is fixed for all elements could thus affect the retrieved abundance. For a fixed OH signal, an underestimation of $R_{\rm P}$ (from assuming constant abundances at all pressures) would lead to overestimation of its abundance. This might explain why we retrieve such an abnormally large OH abundance, which might not be representative of reality.


We still tested a few things to challenge the validity of the signal. 
We first looked for contamination from residual sky OH emission lines, but found no obvious signal following the BERV (either positive or negative). Plus, we already exclude the exposures where \vvv{P} crosses the BERV in all of our analyses. We then verified whether the signal could be an alias originating from cross-correlation between the different molecular absorption spectra, but did not see significant signals around 0\,\kms\ with any other molecular specie that we included in our analyses.

A priori, the presence of OH high in a planet's atmosphere could be due to photochemistry. 
To check, we generated \texttt{VULCAN} models of \waspb\ \citep{VULCAN}\footnote{\texttt{VULCAN} is an open-source photochemical kinetics \texttt{Python} code for exoplanetary atmospheres, in which \texttt{FastChem} is implemented.} that include photo-chemistry, and used the JR retrieved T-P profile. We observed that OH is present in small abundances at $P \gtrsim 1$\,bar ($\simeq 10^{-11}$), slightly decreases until $P \simeq 0.1\,$bar, but re-increases to peak again at values of $\sim 10^{-6}$ around $P \simeq 10^{-7}$\,bar. These amounts are still too low to be detected anywhere in the atmosphere. If we artificially increase the temperature to $\sim 1400$\,K (more representative of what the day side \emph{could} look like), the OH abundance increases to $\sim 10^{-8}$ and $10^{-5}$, at $P \simeq 1$ and $10^{-7}$\,bar, respectively. This abundance could potentially be detected (according to the injection-recovery tests, see appendix \ref{subsec:injec-recov}, Figure~\ref{fig:injec-recov}), but it is still lower than the observed abundance. 
We note that the stellar spectral energy distribution used to compute our VULCAN models might not be representative of the true spectrum of \wasp, as we did not account for the fact that the star is leaving the main sequence. We simply used the same PHOENIX model as the one used to compute the systemic RV above. Also, we assumed a stellar composition for the planetary atmosphere, but another composition could be affected differently. 
The above calculations are rather limited and a more detailed study would be needed to address this possibility.

Other possibilities may explain the OH presence. For instance, if the day side were hot enough to thermally-dissociate some \water, winds could then transport the resulting OH to the terminator region where it would be visible to transit observations.
Also, vertical mixing could bring up OH from deeper and hotter atmospheric layers and make it visible. 
Finally, we cannot exclude the possibility of a spurious signal that happens to match the OH line structures and lands at the correct \vvv{rad}. Overall, strong disequilibrium processes are necessary to maintain the presence of OH.

Obtaining a more representative T-P profile of the day-side of \waspb\ through emission spectroscopy, which will be possible with JWST, could help illuminate this puzzling result. Additional transmission observations at high-resolution would also be beneficial.

\subsubsection{FeH}
\label{subsubsec:FeH}

As briefly mentioned previously, \cite{Skaf2020} reported a tentative detection of FeH which motivated its inclusion in this work. They argued that their signal cannot be caused by clouds, but conclude that it is still possible that they observe another, yet unidentified, opacity source with absorption features resembling those from FeH. \citetalias{Spake2021} also suggest that there may be other unknown and unresolved absorbers between 0.8 and 1.2\,\um. Additionally, \cite{Kesseli2020} found no statistical evidence of FeH in any of the 12 hot-Jupiters they analyzed using high-resolution CARMENES observations. They argued that the previous tentative FeH detections from \cite{Skaf2020} (and others) could be caused by their use of low-resolution spectra, and further insist that it makes it more difficult to distinguish species with overlapping opacities and differentiate them from the continuum opacity. They also show how the FeH condenses at temperatures below 1800\,K, which then makes \waspb\ much too cold to arbor FeH vapor, in an equilibrium standpoint. 
Our SPIRou-only injection-recovery tests indicate that if FeH was present with an abundance down to $\sim 10^{-8}$, we would have easily detected it in our analysis (Appendix \ref{subsec:injec-recov}, Figure~\ref{fig:injec-recov}). From the JR results, we conclude that there is no evidence for FeH in the atmosphere \waspb, at least in an abundance greater than $\sim 10^{\FeHJR}$.

\subsubsection{N-bearing molecules}
\label{subsubsec:HCN}
Our focused searches for HCN and NH$_3$ were unsuccessful. However, when analyzing their posterior distributions in the complete retrieval (Figure~\ref{fig:mcmc_a2}), their upper limits display peaks instead of plateaus like the other non-detections. These peaks are not considered proper detections, but since they are at, or close to, the detection limits for both species, we investigated the question further. Similarly to what we did for OH, we compared the best-fit model against one where we artificially removed HCN and NH$_3$. When excluding both of these molecules, the decrease in CCF S/N, $t$-test and \logl\ are negligible, while testing for a model with only HCN and NH$_3$ yields CCF S/N $= 1.6$ and $t$-value $= 1.7$. 
Given the low significance of this signal, we still consider these non-detections, but do not exclude the possibility that HCN and NH$_3$ are present in the atmosphere of \waspb. Further observations could also help validate this.

\subsection{Comparison between the types of retrievals}
\label{subsec:diff_retrievals}

The retrieved parameters across our three different retrievals are roughly consistent with one another, with, not surprisingly, the best constraints obtained with the JR. 
In all retrievals \water\ is detected, but the LRR and HRR seem to be less constrained and suffer more from degeneracies with \cloud\ and the scattering factor $S$. Likewise, \diox\ is convincingly detected in the LRR, JR, and is consistent with HRR, although it is at the limit of detection, which was expected from our injection recovery tests (Appendix \ref{subsec:injec-recov}, Figure~\ref{fig:injec-recov}). 
From the CO posterior distribution of the LRR, we can see that the lack of independent spectral features from \diox\ allows it to go towards high CO abundances --- a result consistent with the \citetalias{Spake2021} findings. This high CO region is excluded when the SPIRou data is considered thanks to the availability of the 2.3\,\um\ CO band. 
A similar result can be seen for FeH, where the LRR accepts, and even has a peak in the posterior distribution at higher abundance around $\log_{10}$ FeH $\sim $ \FeHLRR\FeHLRRerr. This is, again, compatible within 1$\sigma$ with the findings from \citetalias{Spake2021}: they were able to put a constraint on FeH at $\log_{10}$ FeH $\sim -7.6$, but did not claim a detection. This ``high'' abundance is most likely due to the large range in \cloud\ and $S$ allowed by -- and the limited wavelength coverage of -- the WFC3 and STIS data. 
Looking at the LRR model in Figure~\ref{fig:best-fit-model} (in green), the FeH signal can be seen around 0.9--1\,\um, but no major features are captured by the STIS data points, which permits a certain amount of FeH to be present. Still, the presence of FeH is ruled-out by the SPIRou data.

The \tp\ and $R_{\rm P}$ values of the three retrievals somewhat differ from one to another. LRR seems to prefer higher \tp\ with smaller $R_{\rm P}$, likely coming from the higher values of $S$ that are retrieved. 
The inclusion of the white light transit depth in the HRR helps to better constrain the continuum level, and leads to similar results to those in JR. We observed that by not including the transit depth, of all the retrieved parameters, \tp\ and $R_{\rm P}$ were the most affected, leading to much lower temperatures, but higher radius values. This led to models that had offsets in transit depth, but when they were scaled to match the observed transit depth, the overall structures in the HST/WFC3 data were well reproduced (like the other models in Figure~\ref{fig:best-fit-model}). This means that the HRR is good at extracting the information on the line contrast. 
The retrieved \tp\ from the HRR and JR give T-P profiles that are more in line with the \citetalias{Spake2021} NEMESIS one, while the LRR seems to be more consistent with ATMO ones, even though all are consistent with one another within $\sim 2\,\sigma$.  
As expected, the orbital and rotational parameters (\kp, \vvv{rad}, $\omega$ and $f$; not shown) are all nearly identical for HRR and JR. 

The cloud top pressures inferred from the LRR and HRR retrievals are consistent with one another, which are broader and extends to smaller values (higher-altitude cloud deck), but becomes more constrained in the JR. 
At low-resolution, this arises from the known strong correlations between \cloud\ and hazes, the absolute chemical abundances, the temperature, and also with the reference radius and pressure \citep[e.g.,][]{Benneke2012, Griffith2014, Heng2017_ref_P_and_R}, preventing us from putting tight constraints on the absolute abundances. 
At high-resolution, we benefit from having access to the relative line strengths above the continuum, but the removal of the continuum through data processing also causes additional degeneracies \citep[e.g.,][]{Fisher2019, Brogi2019, Gibson2020}. 
Combining the continuum information from low-resolution with the line shape and contrast at high-resolution (and also relative strength of the broad spectral features; \citealt{Sing2018, Pino2018_combining,Khalafinejad2021, Gibson2022}), while also marginalizing over all affected parameters, allows for much better constraints. 
Overall, even though a combination of slightly higher-clouds and higher-abundances cannot be completely ruled out (down to $\sim 10^{-3}$\,bar in the HRR and JR), the low-clouds and lower-abundances seem to be preferred. 
A cloud level at $\lesssim 10^{-3}\,$bar is ruled out by our JR results.

Concerning the scattering properties, they all seem to agree on non-Rayleigh scattering, with $\gamma$ values larger than $-4$ (although the HRR is insensitive to it). The constraint on the $S$ factor highly benefits from the combination of the low- and high-resolution data. The LRR favors high-$S$ models, while the addition of the high-resolution data prevents it.

Finally, all the C/O posterior distributions are consistent between the retrievals where they all favor sub-solar C/O (Table~\ref{tab:param_MCMC_a2}). 
The LRR allows for super-solar C/O values (beyond the 1$\sigma$ region), but still largely favors the sub-solar C/O region (principal peak at low C/O, but with a long faint tail going up to C/O$=1.0$, not shown). 

For C/H, all retrievals return broad posteriors that are consistent with one another and with stellar compositions. 
For O/H, all retrievals favour stellar to super-stellar abundances, but the LRR and JR values are inconsistent with one another to $1\sigma$.
The overall higher values of C/H and O/H from the LRR seem to arise, again, from the higher values of $S$, the strength of the hazes opacities, while the higher O/H in HRR is driven by its ability to only retrieve super-solar \water; both of these limitations are circumvented in the JR due to the complementarity of the features probed by the two sets of data.

This demonstrates how one can improve the determination of atmospheric parameters by combining low- and high-resolution data sets. The low-resolution allows to get the continuum and broadband information, while the high-resolution allows to probe the line shape and contrast. This grants the ability to lift some degeneracies that would arise from having either one or the other. Future atmospheric analysis will thus benefit from having both types of data.

\subsection{Rotation Kernel and Radial Velocity Offset}
\label{subsec:wind_a2}

As transmission spectroscopy probes only the limb of a planet's atmosphere at the terminator, it is naturally sensitive to global gas movement, such as day-to-night winds or jets and rotation, which result in net blue shifts or broadening of the absorption signal in transit \citep[e.g.][]{Snellen2010, Kempton2012, Flowers2019}. 
Here, when we do not use our custom rotation kernel, we measure a strong blue-shift of \vpeakttestkpo\vpeakttestkpoerr\,\kms\ for \waspb. This shift is seen in each of the three transits, independently of the analysis steps applied. It is hard, however, to conclude whether or not this shift results from global day-to-night winds or something else. 
Such large blue shifts have been observed in several targets. For example, the CCF signal in the \kp\ versus \vvv{rad}\ map of the ultra-hot Jupiter WASP-76\,b \citep{Ehrenreich2020} is highly blue shifted, but this is caused by the signal mostly coming from the trailing (evening) limb of the terminator because of non-uniformity of the atmosphere. \cite{Wardenier2021_wasp76} investigated this further with 3D general circulation models (GCMs) and confirmed that their different models with and without drag also displayed a blue-shifted signal. For some models, it even displays a double peaked CCF. The blue-shift is intensified when they introduce an asymmetry in composition in the terminator limbs (i.e., a depleted abundance in the leading/morning limb compared to the trailing limb). If condensates and/or clouds appear on the night side, the \water\ signal would be muted on the morning limb, while still being relatively strong on the evening limb. We would then expect a faint to invisible signal that is progressively more blue-shifted during ingress, and a globally blue-shifted signal during total transit and egress. Such a "moving" planet trace could also partially broaden the signal. 
We would need a kernel that evolves with time to capture the rotation and dynamics behavior in greater details. 

We attempted to reproduce this behaviour with our kernel of a rotating shell, which can naturally explain the shift. With this choice, we recover tight constraints on the kernel's shape, despite its large range of possibilities (single or double peaked; blue, red or not shifted; symmetric or asymmetric, as shown in Figure~\ref{fig:rot_ker}). 
The JR converged on a rapidly rotating atmosphere/planet ($\omega \simeq 9.1$\,d$^{-1}$) with no additional blue shift (\vvv{rad} $\simeq$ \vpeakJR\vpeakJRerr\,\kms) and a muted right half ($f \simeq 0.49$),\footnote{Note that in our CCF and $t$-test maps (Figures~\ref{fig:CCF_2d_a2}--\ref{fig:ttest_OH}), we see a fainter second peak near \vvv{rad}$\sim 10\,$\kms, such that the overall signal may resemble the two-component, asymmetric rotation kernel that we retrieve here.} which would be consistent with a picture similar to the one painted for WASP-76\,b. 
A signal dominated by the hotter, more inflated, and more blue-shifted trailing limb (from the combination of the planet rotation and and super-rotation from longitudinal advection of the hot-spot with eastward winds, for example), combined with a cloudy leading limb, could explain such a kernel shape. 
Knowing whether and how this could arise for a highly inflated planet like \waspb\ would require a more detailed modelization, but this is beyond the scope of this paper. For instance, 3D GCMs would be useful to see if such a large asymmetry and super-rotation could be explained \citep{Flowers2019, Beltz2021, Wardenier2021_wasp76, Savel2022_wasp76}.

If the shift comes from an asymmetry in temperature and/or composition between the trailing and evening limbs, then our retrieved chemical abundances and temperature profile should be used as estimations only. \cite{MacDonald2020} showed that using 1D atmospheric models could lead to smaller retrieved temperature and smaller temperature gradient than reality, but also to over-estimations of the chemical abundances \citep[also][]{Pluriel2020}. It would thus be highly valuable to apply retrieval techniques using 3D models. 

Alternatively, the blue shift we detect when not using our custom rotational broadening could be caused by the propagation of the error in the transit mid-point, an unnoticed eccentricity, or a combination of a few of all the mentioned hypotheses, as suggested by \citep{Line2021} to explain the noticeable offset in their emission signal of WASP-77\,Ab.

\subsection{Implications for the Formation of WASP-127\,b}
\label{subsec:formation}

In this section, we explore possible formation scenarios under the assumption that the measured elemental abundance ratios reflect the bulk envelope composition of \waspb, and that these are the result of planet formation.



From static disk formation scenarios \citep[e.g.,][]{Oberg2011}, our measured value of C/O $ = $ \CsurOdistnoOH\CsurOdisterrnoOH\ suggests it was done in an oxygen-rich environment, likely due to significant solid accretion. This is the case in \cite{macdonald_metal-rich_2019}, where they put an upper limit of C/O$ < 0.33$ for the exo-Neptune HAT-P-26\,b, with a super-solar metallicity ($\sim 20\times$ solar), and suggested a formation history with significant planetesimal accretion. 
Conversely, our determined stellar C/H and O/H ratios would imply a formation scenario inward of the icelines, where all the volatiles are in gas form, with stellar abundances. 
The combination of these two together is puzzling, as a formation scenario dominated by the accretion of icy solids would tend to elevate the X/H above what we find. 
Alternatively, a gas-dominated accretion within the icelines should result in a solar to super-solar C/O, which is not what we find.
Our observed abundance ratios are thus hard to reconcile with the composition trends predicted by most existing planet formation models that rely on static disks \citep{madhusudhan_exoplanetary_2019}. 

Overall, three scenarios could explain what we observe : 1) either C and O are both stellar-abundant, but some of the C is in invisible phases, 2) C and O directly acquired their respective abundances, or 3) O and C were added back to a metal poor gas, but the accreted material was predominantly O-rich. 
In scenario 1, the C/H could be equal or greater than the O/H, but where a large amount of C-bearing elements would be in forms that are not probed in our transmission spectra, such as condensed hydrocarbons or graphite phases. This would lead to under-estimations of both the C/H and C/O. 
The last two cases would lead to a lower C/H than O/H and a low C/O. We explored the scenarios 2 and 3 more in depth, assuming that the C-bearing molecules are in vapor phase, and investigated composition predictions from studies that include the effects of evolving disk chemistry \citep[e.g.,][]{Eistrup2018, Cevallos2022}.


\cite{Cevallos2022} explored the impact on the planetary atmosphere composition from including the effects of radial pebble drift, gas advection, mass accretion and viscosity in their chemically evolving disks. Pebble accretion is an efficient way to build a core quickly, even with a lower amount of solids, and allows the accretion of substantial amount of gas before the disk disperses \citep{Lambrechts2012_pebble}. 
We get a good match with two of their scenario that includes both advection and pebble drift, with either evolving or fixed mass accretion rate ($10^{-8} \rightarrow 10^{-9}$ or $10^{-9}\,M_{\odot}$\,yr$^{-1}$ , and fixed pebble size (0.3\,cm; their cases d and e), even at times as early as $10^5$\,yrs. In these scenarios, the radial pebble drift puts all the CO in gas-form, while inside $\sim 10$\,AU the \water\ and \diox\ ices are enhanced, in gas or solid form. The inclusion of pebble drift yields ``peaks and valleys" in the gaseous abundance ratio throughout the disk, allows the C/O of both the gas and ice phases to drop to low values, and allows the X/H to locally increase to much higher values, close to what we find for WASP-127\,b. 
A formation close to the \diox\ iceline offers a possible explanation for what we observe. However, it would still require accretion of both gas and ices  from inward migration through the \water\ and \diox\ icelines, with significant envelope-core mixing to closely match our observations \citep[e.g.,][]{Ali-Dib2017_form_loc, madhusudhan_atmospheric_2017, oberg_jupitertextquotesingles_2019}. 
In any case, such a situation would not readily explain the overabundance of \diox\ compared to CO, considering the \diox\ rich ices would be offset by CO rich gas.

The \diox/CO tension could possibly be explained by post-formation evolution. \cite{Fleury2019} showed that irradiation of a H$_2$-CO gas mixture with Ly$\alpha$ promotes very efficient photo-chemistry, leading to CO depletion while enhancing production of \diox\ or \water, depending on the temperature. At temperatures below $\sim 1200\,$K, \diox\ is the main product. 
At higher temperatures, \water\ becomes the dominant product, to the detriment of \diox. Their experiments also show that the production of \diox\ never reaches a plateau, meaning that given the right conditions this process could be kept on for a long time. 
From our results, \waspb's atmosphere is exactly at the right temperature to promote \diox\ production and CO destruction. It is unclear whether photo-chemistry could deplete the upper atmosphere of WASP-127\,b of CO and bring the results of the scenario (d) of \citet{Cevallos2022} in line with what we observe. 

Another possibility is for the disparity in CO and \diox\ to be already present at the time of formation, and somehow has survived until today. 
This is possible when considering the freeze-out of CO in \diox\ ices as presented in \cite{Eistrup2018}. They show that after 5\,Myr, photo-chemistry in the disk leads to a large drop in CO gas and an increase in frozen \diox\ beyond $\sim 10\,$AU (see their Figure~7c). This comes from the collision of CO gas with solid grains, and the subsequent fast reaction with OH to produce \diox\ on the grain surface\footnote{The OH radicals can be produced by the dissociation of \water\ by cosmic rays, which also diminish the local \water\ abundance, but in this case it would be produced via O$_2$ freeze-out. Beyond the O$_2$ iceline, the O$_2$ freezes-out, is further hydrogenated on grains, is then dissociated into OH ice, and finally reacts with the CO to form \diox. }. 

This late formation could be a plausible scenario given that the low-metallicity of \wasp\ implies a lower density of solids in the protoplanetary disk --- meaning that the initial planet core growth is more difficult and thus takes longer in the core accretion scenario \citep[][]{Alessi2018}. 
By the time it was ready to accrete gas in a runaway fashion, the disk could have been already partly dissipated (as demonstrated in \citealt{Eistrup2018}). Such a scenario could furthermore explain why the growth of \waspb\ was truncated at a sub-Saturn mass instead of continuing to a super-Jupiter size --- the requisite materials were simply no longer present in the disk to facilitate this runaway growth \citep[][]{pollack_formation_1996}. But compared to their computed values, additional accretion of \water\ and \diox\ ices (both still highly abundant in the disk, and available as building blocks for the core) with mixing in the envelope would help increase the O/H, while lowering the gaseous C/O to our observed values.

\section{Concluding remarks}
\label{sec:conclusion_a2}

This work presented the first high-resolution near-infrared transmission spectrum of \waspb. We combined our high-resolution SPIRou data with existing HST and \emph{Spitzer} data from \cite{Spake2021} to perform a joint high and low-resolution MCMC retrievals and place improved constraints on the atmospheric parameters of \waspb.

Using the SPIRou data only, we clearly recovered the \water\ detection at a level of $\sim 5 \sigma$ and found a tentative signal of OH (\OHttest$\sigma$). The presence of OH would be surprising as it is not expected to exist at the temperature of the planet's terminator probed by our observations. If the signal is real, OH must come from other parts of the atmosphere and be brought to the terminator, or it could arise from photo-chemistry. No other chemical element was retrieved with statistical significance when using only the SPIRou data. 
Our injection-recovery tests showed that we are nonetheless sufficiently sensitive to most of the tested elements to put useful upper limits on their abundance in the atmosphere of \waspb. 
We performed a series of retrievals, at low- and/or high-resolution, to better compare the different results, and the limitations of each retrieval type. As expected, the parameters are better constrained with the joint retrieval, combining both low- and high-resolution observations, and our main results are based on that. 

One important result of this work is the non-detection of CO, with an upper limit of $\log_{10}$CO $ < $\COJR, which excludes the chemical equilibrium scenario from \citetalias{Spake2021}. This indicates that the strong signal from the 4.5\,\um\ IRAC2 band mainly comes from \diox\ absorption, the abundance of which we constrain to be $\log_{10}$\diox $ = $ \dioxJR\dioxJRerr. The presence of \diox\ and lack of CO likely mean that non-equilibrium processes are at play in \waspb's atmosphere, and how this imbalance is maintained remains to be explained. 

Our retrieved abundances yield severely sub-solar C/O (C/O $ = $ \CsurOdistnoOH\CsurOdisterrnoOH), resulting from C/H and O/H that are still both consistent with stellar abundance. 
Such disparity between the elemental abundances is unexpected when considering the more standard formation scenarios, assuming that all the major C-bearing elements are in gas phase rather than hidden in a condensed form. 
However, when considering the effects of radial pebble drift, gas advection, mass accretion and viscosity (such as was done by \citealt{Cevallos2022}), we find that a core accretion scenario, with accretion of both gas and ices close between the \water and, close to, the \diox\ ice line could potentially produce an atmosphere with the abundances we observed. It would still require envelope-core mixing, further migration and enrichment through the ice lines, and subsequent CO depletion. 
Alternatively, a late/slow formation scenario beyond 10\,AU within a chemically evolving disk, where the bulk of the accretion of the gas occurred between $\sim 5$ and 7\,Myrs, offers an interesting formation avenue. In such a scenario, photo-chemistry on \diox\ grains in the disk causes CO gas depletion that can explain the discrepancy between our CO and \diox\ abundances, but need to have survived until now. The expected composition in such models is consistent with our observations when considering further accretion of O-rich material, such as \water\ or \diox\ ices (from crossing their ice lines), that would be vaporised in the atmosphere to increase the O/H and lower the C/O.

We showed that by doing joint retrievals, we are able to lift degeneracies that affect transmission spectroscopy at high or low-resolution, independently.
Indeed, our joint retrieval results still show some correlation between abundances, $T_{\rm P}$, \cloud, and $R_{\rm P}$, but is much more constraining than with only high- or low-resolution. It favors a relatively clear atmosphere, with $R_{\rm P}$ being consistent with the value from \cite{Seidel2020_wasp127}. 

From the \kp\ versus \vvv{rad} $t$-test maps, we detect the planet signal at a large blue shift of \vvv{rad} of $\sim $\vpeakttestkpo\,\kms. Our retrieved values for the frequency of our solid rotation kernel and the relative strength between the two hemispheres suggest that this shift is caused by an evening limb dominated signal, with a cloudy/muted morning terminator, combined with super-rotation, which highly broadens the signal.  
A more detailed 3D global circulation modeling of \waspb's atmosphere would be highly beneficial to explain these large blue shift and broadening, and better represent the overall dynamics at play.

Given that many things remain unclear about this fascinating planet, \waspb\ is most certainly a prime target for future observations, from the ground at high-resolution and at lower resolution with JWST. More high-resolution observations would be useful, for instance, to confirm or rule out our tentative OH detection. Also, observations of the day side, through emission spectroscopy, would be particularly useful to constrain the T-P profile and chemistry of the hot hemisphere, as well as inform global circulation models. A more in-depth study of the stellar parameters is also needed to better anchor and compare to the current observations. The fact that the star is leaving the main-sequence could be a factor in explaining what we see here.

\section*{Acknowledgements}

We thank the referee Paolo Giacobbe for the careful reading, the feedback, and suggestions that really helped to improve the quality of this manuscript.
The authors acknowledge financial support for this research from the Natural Science and Engineering Research Council of Canada (NSERC), the Institute for Research on Exoplanets (iREx) and the University of Montreal.
These results are based on observations obtained at the Canada-France-Hawaii Telescope (CFHT) which is operated from the summit of Maunakea by the National Research Council of Canada, the Institut National des Sciences de l'Univers of the Centre National de la Recherche Scientifique of France, and the University of Hawaii. The observations at the Canada-France-Hawaii Telescope were performed with care and respect from the summit of Maunakea which is a significant cultural and historic site. 
SP and ADB acknowledge funding from the Technologies for Exo-Planetary Science (TEPS) CREATE program. 
FD thanks the CNRS/INSU Programme National de Planétologie (PNP) and Programme National de Physique Stellaire (PNPS) for funding support.
XD acknowledges funding from the ANR of France under contract number ANR18CE310019 (SPlaSH) and in the framework of the Investissements d'Avenir program (ANR-15-IDEX-02), through the funding of the “Origin of Life" project of the Grenoble-Alpes University.
JFD acknowledges funding from the European Research Council (ERC) under the H2020 research \& innovation programme (grant agreement \#740651 NewWorlds). 
EG acknowledges support from NASA awards 80NSSC20K0957 and 80NSSC20K0251 (Exoplanets Research Program)
TF is supported by the ANR in the framework of the Investissements d’Avenir program (ANR-15-IDEX-02), through the funding of the “Origin of Life” project of the Grenoble-Alpes University
J.H.C.M. is supported in the form of a work contract funded by Fundação para a Ciência e Tecnologia (FCT) with the reference DL 57/2016/CP1364/CT0007; and also supported from FCT through national funds and by FEDER-Fundo Europeu de Desenvolvimento Regional through COMPETE2020-Programa Operacional Competitividade e Internacionalização for these grants UIDB/04434/2020 \& UIDP/04434/2020, PTDC/FIS-AST/32113/2017 \& POCI-01-0145-FEDER-032113, PTDC/FIS-AST/28953/2017 \& POCI-01-0145-FEDER-028953, PTDC/FIS-AST/29942/2017. J.H.C.M. is also supported by supported by FCT - Funda\c{c}\~ao para a Ci\^encia e a Tecnologia through national funds and by FEDER through COMPETE2020 - Programa Operacional Competitividade e Internacionaliza\c{c}\~ao by these grants: UID/FIS/04434/2019; UIDB/04434/2020; UIDP/04434/2020; PTDC/FIS-AST/32113/2017 \& POCI-01-0145-FEDER-032113; PTDC/FIS-AST/28953/2017 \& POCI-01-0145-FEDER-028953; PTDC/FIS-AST/28987/2017 \& POCI-01-0145-FEDER-028987.

We appreciate all the work done by the developers of the \texttt{NumPy},  \texttt{SciPy}, \texttt{Matplotlib}, and \texttt{Astropy} packages, which represent truly amazing tools and are at the base of this work \citep{Jones2001_SciPy, Hunter2007_matplotlib, Astropy2013}. We are also grateful to the developers of petitRADTRANS for making their code publicly available, which allowed to easily and quickly produce atmospheric models independently for our retrievals.

\section*{Data Availability}

All the publicly released SPIRou data, raw and reduced, can be found at the \href{https://www.cadc-ccda.hia-iha.nrc-cnrc.gc.ca/en/cfht/}{CFHT Science Archive}.



\bibliographystyle{mnras}


\bibliography{ref} 






\clearpage
\appendix

\section{Development of the Rotation Kernel}
\label{an:rot_ker}

\begin{figure}
\includegraphics[width=\linewidth]{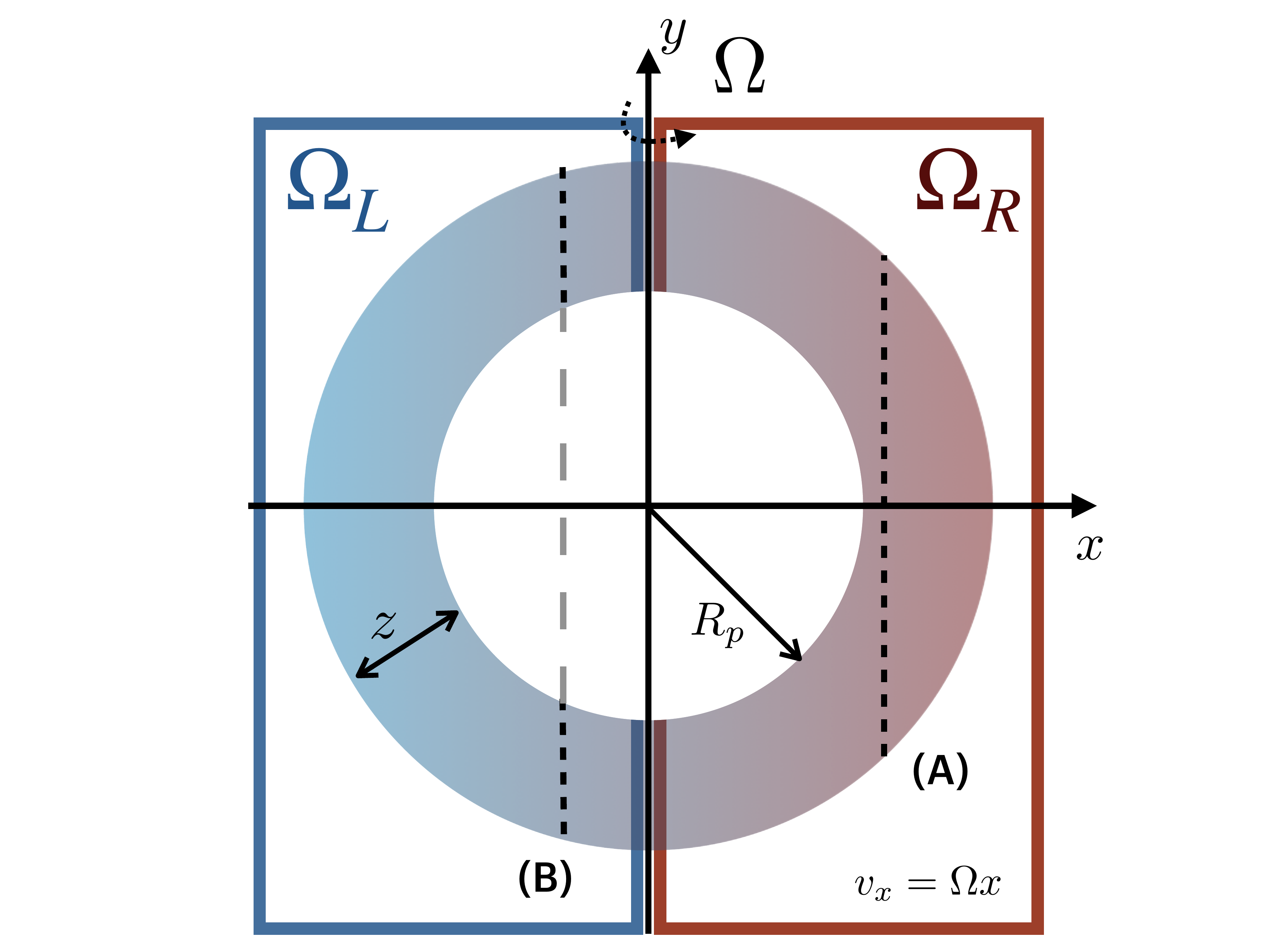}
\includegraphics[width=\linewidth]{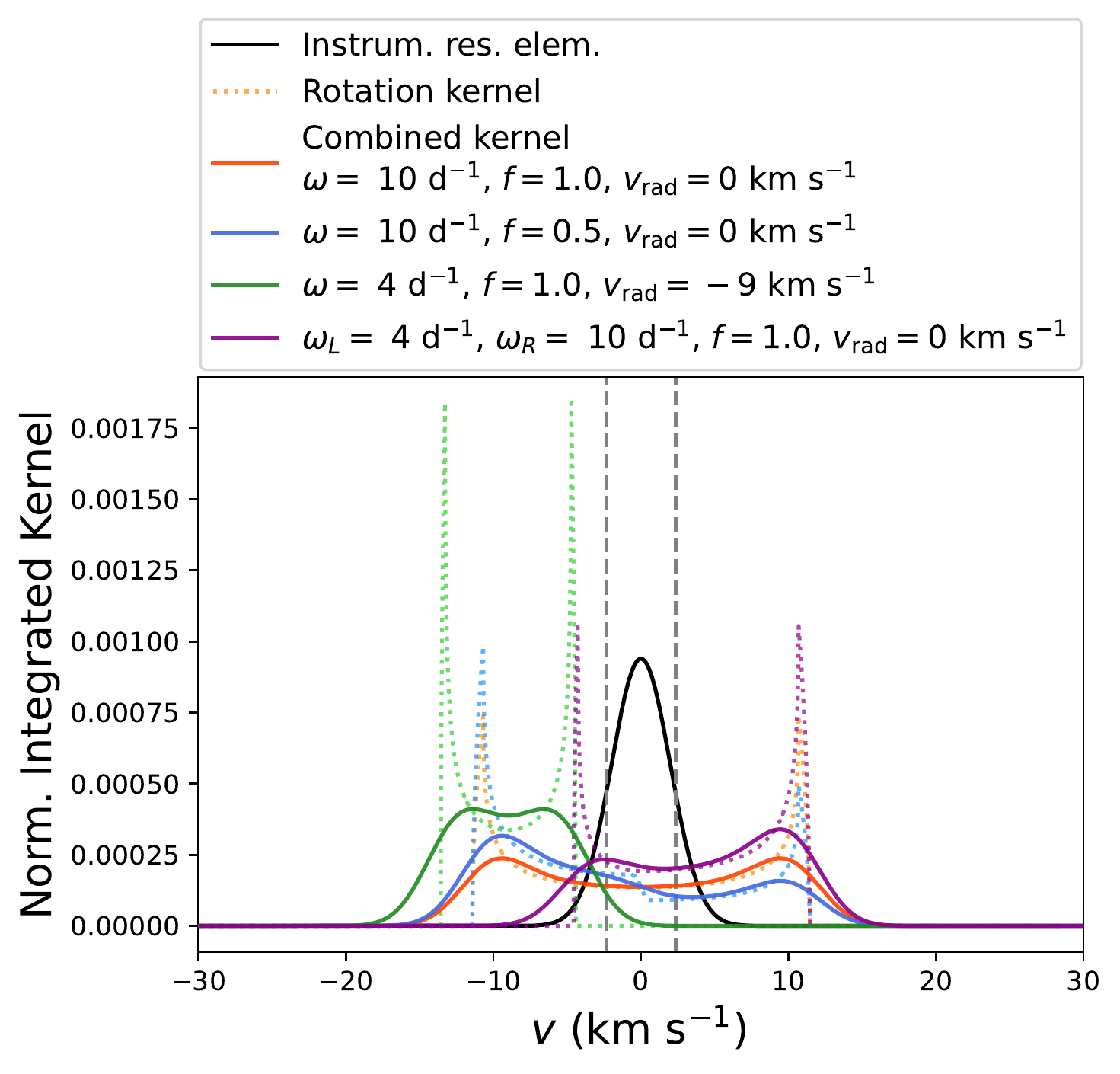}
\vspace{-0.5cm}
\caption{(Top) Qualitative front-view representation of the projected line-of-sight velocities as seen by an observer during transit for a given angular speed $\Omega$. The red parts will appear red-shifted and the blue parts blue-shifted. In the solid rotation case, only $\Omega$ (in black) is relevant and dictates the equatorial velocity in both hemispheres, while in the double rotation case, $\Omega_L$ and $\Omega_R$ translate to the independent left (blue region) and right (red region) equatorial velocities, respectively. In this work, we fix $\Omega_L = \Omega_R = \omega$. 
(Bottom) Examples of rotation kernel with different $\omega$, $f$, and \vvv{rad}. The instrument profile is shown in black, the dotted curves show transit rotation kernels, and the colored curves show the combined kernels (the rotation kernel convolved with the instrument profile).
}
\label{fig:rot_ker}
\end{figure}

We used a simple rigid shell representation to compute the rotation kernel of the atmosphere in transit. 
For a distance $x$ perpendicularly away from the star rotation axis, the Doppler shift $\Delta v$, relative to $x=0$, is constant for all values of $y$ in the direction parallel to the rotation axis and is given by :
\begin{equation}
    \frac{\Delta \lambda}{\lambda} = \frac{\Delta v}{c} = \frac{\Omega x}{c} \,.
\end{equation}
In other words, all points on the chord at $x$ are subject to the same Doppler shift (see Figure~\ref{fig:rot_ker}, top panel).
The kernel shape will be dictated by the ratio $I_{\lambda}(\Delta v)/I_{\lambda}(0)$, the intensity of a transmission (or emission) line at Doppler shift $\Delta v$ compared with its intensity at the line center ($v_0=0$). This, in turn, is equal to the ratio between the height of the constant chord at $x$ and its height at $x=0$, equal to twice the height of the atmosphere (2$z$). Here, the thickness of the atmosphere $z$ is approximated by $z = 5H$, where $H$ is the atmospheric scale height. 
The kernel shape is then given by: 
\begin{equation}
    \begin{aligned}
        \frac{I_{\lambda}(\Delta v)}{I_{\lambda}(0)} &= 2\frac{\sqrt{(R_p+z)^2-x^2}}{2z} \\
        &= \frac{\sqrt{(R_p+z)^2-(\Delta v/\Omega)^2}}{z} \,,
    \end{aligned}
\end{equation}
if $R_p \leq |x| < R_p+z$ (see Figure~\ref{fig:rot_ker} , top panel (A)), and by:
\begin{equation}
    \begin{aligned}
        \frac{I_{\lambda}(\Delta v)}{I_{\lambda}(0)} &= 2\frac{\sqrt{(R_p+z)^2-x^2}-\sqrt{R_p^2-x^2}}{2z} \\
        &= \frac{\sqrt{(R_p+z)^2-(\Delta v/\Omega)^2}-\sqrt{R_p^2-(\Delta v/\Omega)^2}}{z} \,,
    \end{aligned}
\end{equation}
if $|x| < R_p$ (Figure~\ref{fig:rot_ker}, top panel  (B)). We allow for the kernels to have different $\Omega_L$ and $\Omega_R$, i.e., for the two hemispheres have different rotation angular frequencies, again, to capture possible different integrated velocities (see Figure~\ref{fig:rot_ker}, bottom panel, magenta curve). Our tests led to values that were very similar for $\Omega_L$ and $\Omega_R$. We thus used a single rotation rate and fixed $\Omega$ = $\omega$, the solid rotation angular frequency. 
Additionally, to simulate the effect of each hemisphere having a different cloud fraction or temperature, we added a scaling factor $f$ defined as the ratio between the right ($\Delta v > 0$) and the left ($\Delta v < 0$) parts of the kernel. Its effect on the kernel translates into a multiplicative step function. 
Some examples are shown in Figure~\ref{fig:rot_ker} (bottom) to illustrate the effect that each parameter ($\omega$, $f$, and \vvv{rad}) has on the kernel. Small $\omega$ values can lead to single-peaked kernels while large values are double-peaked, $f \neq 1$ leads to asymmetries, and \vvv{rad} moves the kernel as a whole. Naturally, this simple kernel does not include the 3D effects that would affect a photon passing through the different layers of the atmosphere, but due to the thinness of it, compared to $R_{\rm P}$, these effects are likely negligible, especially since \cite{Beltz2021} showed that Doppler effects were only of secondary orders compared to the temperature structure on the shape of the planetary signal. 

\section{Injection-Recovery tests}
\label{subsec:injec-recov}

To assess the detection significance level of various molecules in our data, we performed an injection-recovery test on the same molecules included in the high-resolution detection analysis from Section~\ref{sec:results_a2}. 

We produced synthetic transmission spectra including all the above molecules at their chemical equilibrium abundances, along with hydrogen and helium. We computed mean chemical equilibrium abundances of each element with \texttt{FastChem} \citep{FastChem}, using \citeauthor{Spake2021}'s $Z_{\rm trace}$, O, C and Na retrieved abundance and T-P profile (that we found equivalent to setting $T_{\rm eq} = 1150$\,K using our Guillot profile with fixed $\kappa_{\rm IR}$, $\gamma$ and $T_{\rm int}$)\footnote{Because of how we computed the chemical equilibrium abundances and their respective 1\,$\sigma$ error bars (explained above), there might be slight discrepancies between \citetalias{Spake2021}'s abundances and ours, but the relative abundances stays the same, and it should not affect our conclusions.}. 
Next, to test the detectability of a specific molecule, only its VMR was varied, with values ranging from $10^{-9}$ to $10^{-2}$, while all the other molecules VMRs were held fixed at their equilibrium VMR value. 
For all models, we tested two cases: one with a clear atmosphere (\cloud $ = 10^2$\,bar, and no haze), and one with the scattering conditions from \citetalias{Spake2021} (i.e., no grey clouds, but with $\gamma = -1.50$ and $S = e^{2.68} \simeq 14.6$, the scattering parameters from equation~\ref{eq:scat}); no rotational broadening was applied\footnote{We tested the impact of the rotational broadening on these limits, and found that the abundances were affected by less than an order of magnitude, meaning that our conclusions, especially for CO, still stand.}.

\begin{figure}
\includegraphics[width=\linewidth]{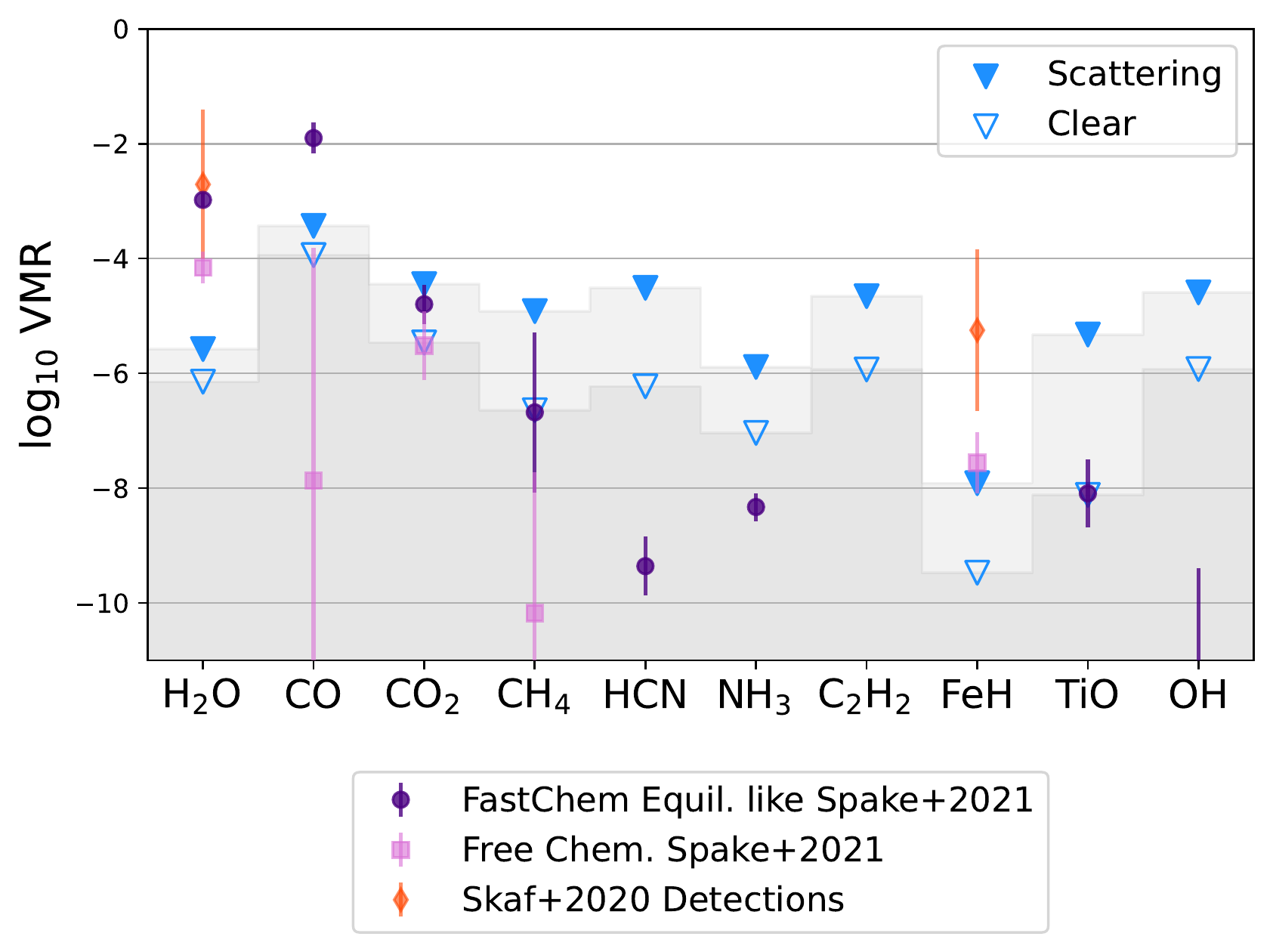}
\caption{Detection limits of the abundances of various molecules in modeled atmospheres with and without hazes (filled and empty triangles), for the combined SPIRou transits and using the $\Delta$BIC significance metric. The abundances in the shaded grey regions are not expected to be detected with our SPIRou data. 
The limits are compared to the results from \citetalias{Spake2021}: for the chemical equilibrium (dark purple circles) and free chemistry (light pink squares) scenarios; as well as the abundances of detected molecules in \citealt{Skaf2020} (orange diamonds).  
When the triangles are below the purple/pink/orange symbols, we can expect to detect the specific molecules, if it indeed follows the predictions from these scenarios. 
With the SPIRou data, we are thus very sensitive to \water\ and FeH, we should be able to detect CO in the equilibrium chemistry case, and are at the limit of detection for \diox.
}
\label{fig:injec-recov}
\end{figure}

We injected the models at \kp\ and a velocity of \vvv{rad} = 35.0\,\kms, away from the real signal to avoid contamination, in the three transits. 
For each molecule and VMR tested, we computed the \logl\ first with the model that was injected, at \kp\ and \vvv{rad}, and second with a model including the same abundances except for the tested molecule, which was set to zero. 
This way, we can see how the data is able to differentiate between models with and without the tested molecule, and at what level it is preferred.
The $\Delta$\,BIC was then obtained by comparing these two values. From the $\Delta$\,BIC values at all VMRs tested, we used linear interpolation to determine the VMR value that would yield a $\Delta$\,BIC of 10, which was taken as our detection limit. 
The results are shown in Figure~\ref{fig:injec-recov}. 
Our limits are compared with the retrieved abundances from the two scenarios in \citetalias{Spake2021} (chemical equilibrium and the free chemistry). We also added the retrieved abundances of the detected molecules from \cite{Skaf2020} for comparison. Although they are not shown here, the $t$-test metrics yielded similar detection limits, while the CCF yielded slightly worse limits.

The addition of the scattering hazes worsens the detection limits (as expected), sometimes by more than 2 orders of magnitude (the worst case being for TiO, where the bulk of its signal comes from the blue side of the SPIRou wavelength coverage, most affected by the hazes). 
In any case, the detectable CO abundance is lower than the expected chemical equilibrium scenario, meaning that we should be able to confirm or rule it out using our data. We should also be able to easily detect \water, and could detect equilibrium abundances of \diox\ and would be at the limit of detection for TiO in a clear atmosphere. For the free chemistry values of \citetalias{Spake2021}, however, we can only expect to detect \water\ and FeH (or any other molecules with abundances above these limits): the \diox\ abundance is just under the limit of detection, meaning that it could be hard to detect it with high significance when working only with our SPIRou data. 


\bsp	
\label{lastpage}
\end{document}